\documentclass{emulateapj}
\usepackage{graphicx}
\usepackage{subfigure}
\usepackage{amssymb}
\usepackage{natbib}
\usepackage{xcolor}
\usepackage{subfigure}
\usepackage{ulem}
\usepackage{multirow}
\usepackage{hyperref}

\hypersetup{
    colorlinks=true,
    linkcolor=red,
    filecolor=magenta,
    urlcolor=cyan,
    bookmarks=false,
}

\def\xmm{{\sl XMM-Newton}}
\def\chandra{{\sl Chandra}}

\def\ergs{\hbox{erg s$^{-1}$ }}

\def\cmcu{\hbox{cm$^{-3}$}}
\def\kmps{\hbox{km $\rm{s^{-1}}$}}

\def\fexx{Fe~{\sc xx}}
\def\fexix{Fe~{\sc xix}}
\def\fexviii{Fe~{\sc xviii}}
\def\fexvii{Fe~{\sc xvii}}
\def\fexvi{Fe~{\sc xvi}}

\def\nvi{N~{\sc vi}}
\def\nvii{N~{\sc vii}}

\def\cvi{C~{\sc vi}}

\def\neix{Ne~{\sc ix}}
\def\nex{Ne~{\sc x}}

\def\oviii{O~{\sc viii}}
\def\ovii{O~{\sc vii}}

\shorttitle{Evidences for a past AGN half a million years ago}
\shortauthors{Zhang et al.}

\begin{document}

\title{XMM-Newton RGS spectroscopy of the M31 bulge. \\I: Evidences for a past AGN half a million years ago}


\author{Shuinai Zhang\altaffilmark{1,2}, Q. Daniel Wang\altaffilmark{3}, Adam R. Foster\altaffilmark{4}, Wei Sun\altaffilmark{1,2}, Zhiyuan Li\altaffilmark{5}, and Li Ji\altaffilmark{1,2}}

\email{snzhang@pmo.ac.cn}


\altaffiltext{1}{Purple Mountain Observatory, Chinese Academy of Sciences, China}
\altaffiltext{2}{Key Laboratory of Dark Matter and Space Astronomy, CAS, China}
\altaffiltext{3}{Department of Astronomy, UMASS, USA}
\altaffiltext{4}{Harvard-Smithsonian Center for Astrophysics, USA}
\altaffiltext{5}{School of Astronomy and Space Science, Nanjing University, China}


\begin{abstract}
Existing analysis based on spectra from the Reflection Grating Spectrometer (RGS) onboard \xmm\ already shows that the G-ratio of the \ovii~He$\alpha$ triplet in the inner bulge of M31 is too high to be consistent with a pure optically thin thermal plasma in collisional ionization equilibrium (CIE).
Different processes that may affect properties of diffuse hot plasma were proposed, such as resonance scattering (RS) and charge exchange (CX) with cold gas.
To determine which physical process(es) may be responsible for this inconsistency, we present a systematic spectroscopic analysis based on 0.8 Ms \xmm/RGS data, together with complementary \chandra/ACIS-S images.
The combination of these data enables us to reveal multiple non-CIE spectroscopic diagnostics, including but not limited to the large G-ratios of He$\alpha$ triplets (\ovii, \nvi, and \neix) and  the high Lyman series line ratios (\oviii~Ly$\beta$/Ly$\alpha$ and Ly$\gamma$/Ly$\alpha$, and \nvii~Ly$\beta$/Ly$\alpha$), which are not expected for a CIE plasma, and the high iron line ratios (\fexviii~14.2 \AA/\fexvii~17 \AA\ and \fexvii~15 \AA/17 \AA), which suggest much higher temperatures than other line ratios, as well as their spatial variations.
Neither CX nor RS explains all these spectroscopic diagnostics satisfactorily.
Alternatively, we find that an active galactic nucleus (AGN) relic scenario provides a plausible explanation for virtually all the signatures.
We estimate that an AGN was present at the center of M31 about half a million years ago and that the initial ionization parameter $\xi$ of the relic plasma is in the range of 3--4.
\end{abstract}

\maketitle

\section{Introduction}
The nature of diffuse soft X-ray line emission from nearby galaxies remains largely uncertain.
It has been commonly assumed that the emission arises from optically thin thermal plasma in collisional ionization equilibrium (CIE).
Based on this assumption, one may then estimate the mass, energy, and chemical contents of the diffuse plasma.
However, clear evidence for deviations of X-ray emitting plasma from CIE states has been shown in \xmm/RGS spectra of a dozen nearby non-active galactic nucleus (AGN) galaxies \citep{wang12}, where the resonance line ({\it r} line) of the \ovii~He$\alpha$ triplet is weaker than the forbidden line ({\it f} line), or more quantitatively the G-ratio$>1.4$, which is not expected from a CIE plasma (defined as the ratio of  the {\it f} line plus intercombination line to the {\it r} line: $(f+i)/r$).
Without understanding the mechanisms leading to the contamination of the X-ray emission and their spectroscopic and spatial distribution imprints, one cannot reliably estimate the diffuse plasma properties.

Because of its proximity, M31 provides an ideal laboratory for an in-depth X-ray spectroscopic study (distance $\sim$780 kpc, and 1$'\,\sim$0.23 kpc from NED\footnote{The NASA/IPAC Extragalactic Database (http://ned.ipac.caltech.edu).}).
The bulge region of the galaxy is quiescent, showing little cool gas \citep{li09, viaene14, melchior17}, no recent massive star formation \citep[e.g.][]{azimlu11, kang12, dong18}, and no current AGN activity \citep{li11}.
However, the \ovii~G-ratio of its hot gas was also reported larger than 1.46 \citep{liu10}.
It could be due to the weakening of the {\it r} line if resonant scattering (RS) process is important \citep{chen18}, or to the strengthening of the {\it f} line, which may result from such a process as charge exchange (CX) or recombination.

RS of the \ovii~{\it r} line photons could be significant in the M31 bulge because O$^{6+}$ ions are expected to be abundant in disk galaxies where the characteristic temperature of diffuse hot plasma is $\sim10^{6.3}$ K \citep[e.g.][]{wang10}.
A RS model based on Monte Carlo simulation has been applied to the RGS spectrum of its \ovii+\oviii~complex in the 18--23 \AA\ wavelength range. 
The best-fit result suggests that a 0.2 keV plasma with a half solar oxygen abundance \citep{grevesse98} explains both the G-ratio and the broadening profiles of oxygen lines \citep{chen18}.

Alternatively, the large G-ratio may also be explained by CX, which occurs at the interface between hot plasma and neutral gas.
In a CX event, a hot ion steals an electron from a neutral atom (most likely hydrogen and helium) to an excited energy level.
As this electron cascades to lower energy levels, it produces photons that lead to an enhanced {\it f} line.
CX may explain the enhanced \ovii~{\it f} line observed in M82 \citep{ranalli08, liu10}, accounting for $\sim$25\% of the observed flux from the central outflow region, or $\sim$50\% flux from the `Cap' region that is 11.6 kpc away from the galactic disk \citep{zhang14}.
The interpretation was also proposed for the central region of M31 \citep{liu10}.

In this work, we examine these two very distinct scenarios, and will further explore the possibility that the large G-ratio is the signature of an AGN relic.
Indeed, such interpretation has been employed to explain phenomena observed in other wavelength bands and to infer recent AGN activity (see Sec. 5 for details).
While it is difficult to trace the history of currently dormant supermassive black holes in other bands, signatures of AGN relic can be readily observed in the X-ray band for a more prolonged period, because the recombination timescale in diffuse hot gas can be substantially long.

Although the three processes mentioned previously can generate similar high \ovii~G-ratios, they are distinguishable by other spectroscopic and/or spatial diagnostics.
Thanks to the high spectral resolution of the RGS spectrum and the high spatial resolution of the \chandra/ACIS-S image, as well as the newest atomic database (AtomDB v3.0.9\footnote{http://www.atomdb.org}), we can now examine multiple non-CIE diagnostics to determine the dominant process that conduces to the high G-ratio in the M31 bulge.

The paper is organized as follows.
The observations and reduction techniques are described in Sec. 2.
The spectroscopic analysis on the RGS data is presented in Sec. 3.
We examine the CX and RS scenarios in Sec. 4 and explore the AGN relic scenario in Sec. 5.
Our results and conclusions are summarized in Sec. 6.

Throughout the paper, we use the chi-square statistic in the fitting, and quoted errors refer to 90\% confidence level for one free parameter case.

\section{ACIS-S Images and RGS Spectra of M31 Bulge}

\begin{figure*}[htbp] 
 \centering
        \includegraphics[angle=0,width=0.45\textwidth]{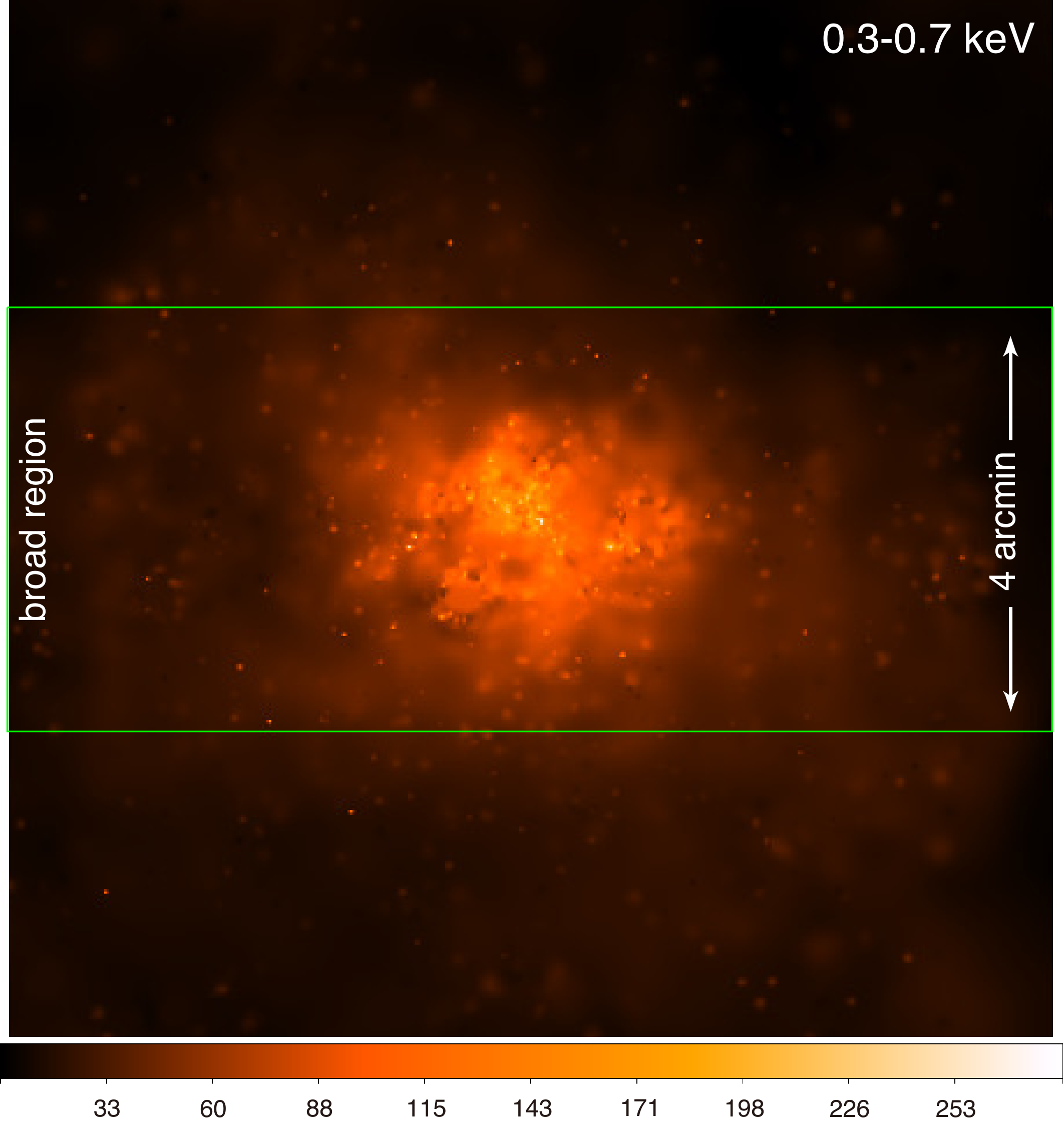} 
       \includegraphics[angle=0,width=0.45\textwidth]{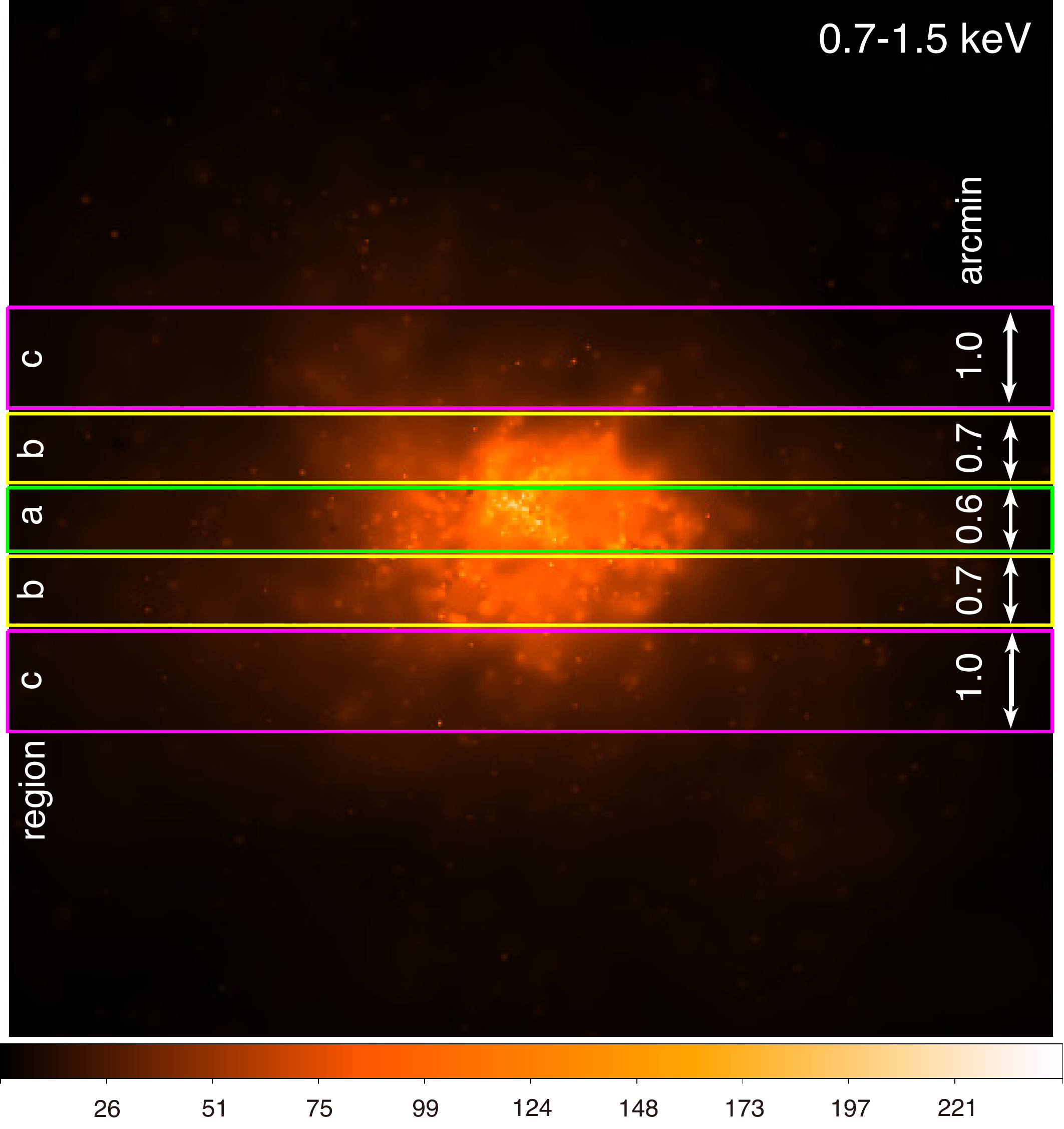} 
 \caption{Intensity maps of the diffuse hot plasma deduced from the \chandra/ACIS-S observations in the 0.3--0.7 and 0.7--1.5 keV bands. 
The spatial coverage of the ``broad region'' is illustrated in the left image with the green box, and that of the ``region a to c'' is represented in the right image with green, yellow, and magenta boxes, respectively.}
\label{fig:chandra} 
\end{figure*}

\begin{figure}[htbp] 
 \centering
\subfigure[]{
\begin{minipage}[t]{3.3in}
\centering
       \includegraphics[angle=0,width=3.2in]{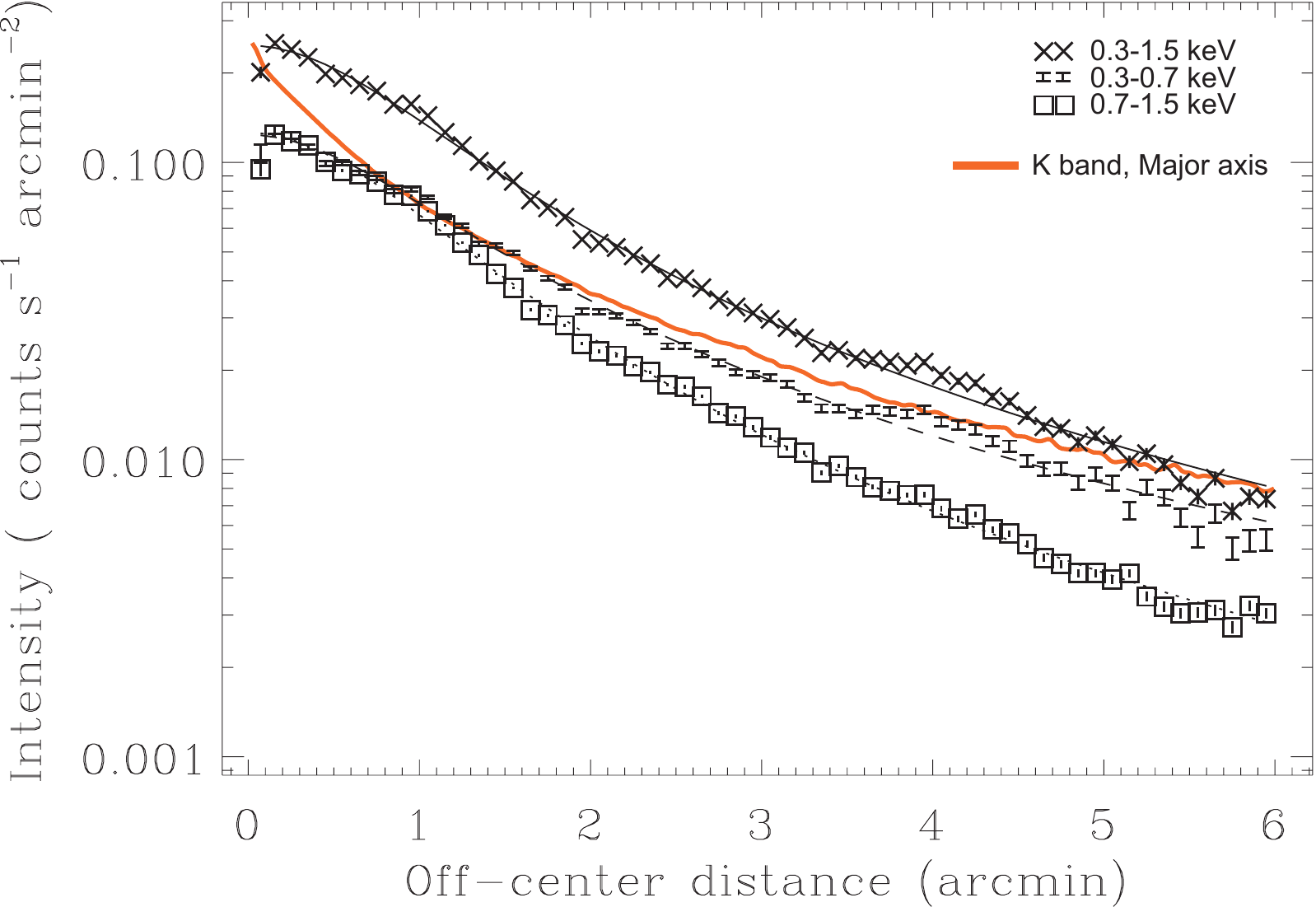} 
\end{minipage}%
}%

\subfigure[]{
\begin{minipage}[t]{3.4in}
\centering
       \includegraphics[angle=0,width=3.2in]{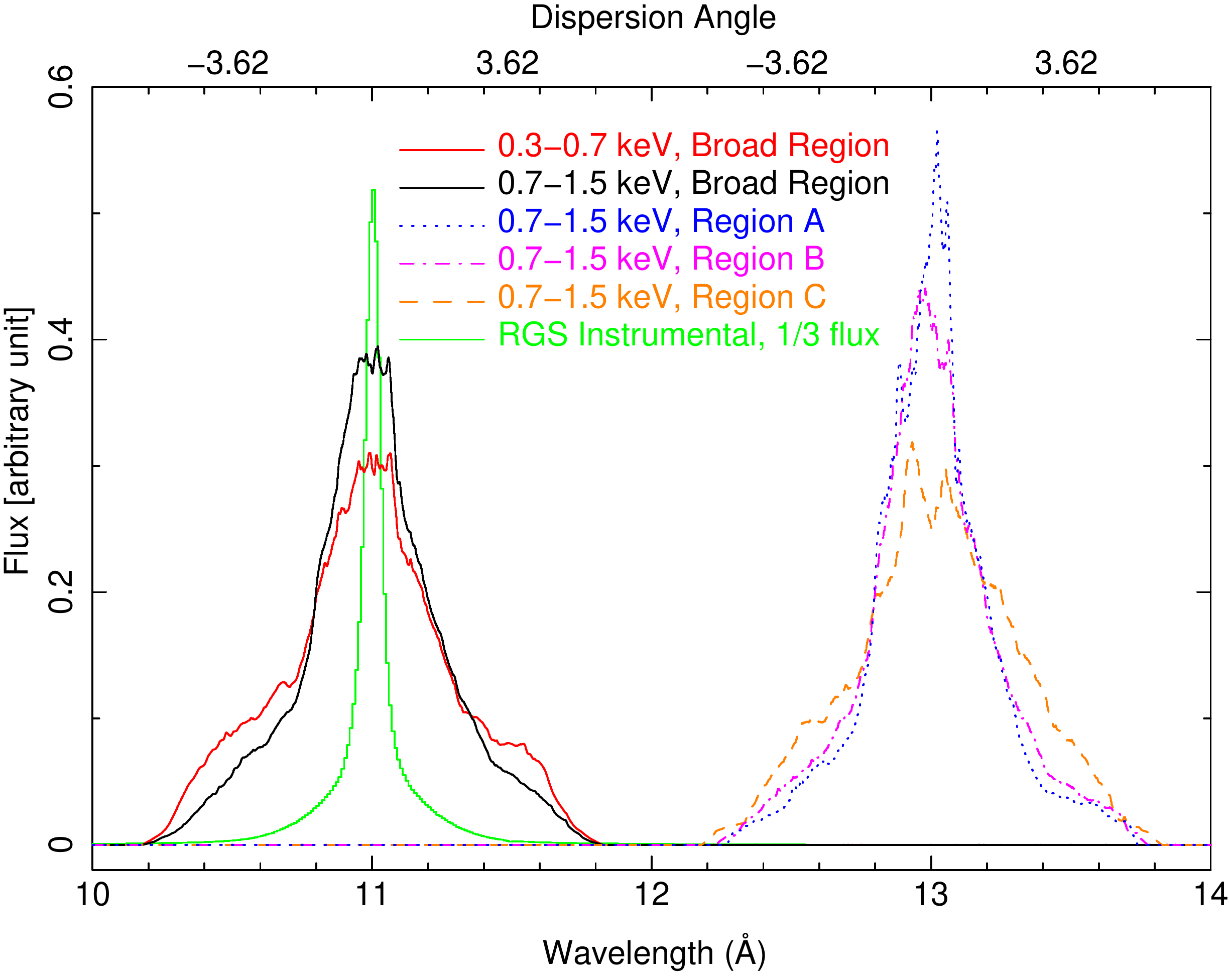} 
\end{minipage}%
}%
 \caption{(a) Surface brightness profiles of the hot plasma in the bands of 0.3--0.7 keV, 0.7--1.5 keV, and 0.3--1.5 keV. 
The orange curve shows the $K_s$-band intensity profile along the major axis of M31 with a normalization of $\rm 2\times 10^{-4}cts\,s^{-1}arcmin^{-2}/MJy\,sr^{-1}$.
(b) Illustrations of line broadening profiles for different combinations of energy band and spectral extraction region, as noted in the plot, compared with the RGS instrumental broadening (solid green curve; presented only at one-third of the flux).
}
\label{fig:bulge} 
\end{figure}

\subsection{\chandra/ACIS-S Images and $\beta$-model}
We reduce \chandra~ACIS-S data to study the spatial distribution of the X-ray emission, which also influences the line broadening in an RGS spectrum.
The on-axis back-illuminated S3 chip of ACIS-S is sensitive to soft X-ray emission with low energy limit pushed down to 0.3 keV, compatible with the RGS band.
We utilized all available ACIS-S observations with a total exposure of 380 ks, as listed in Table~\ref{tab:chandralog} in Appendix A.
The procedure followed to generate the intensity maps of diffuse hot plasma is the same as in \citet{li07}.
Briefly, we detect and excise bright point sources, and subtract the contribution from unresolved faint stellar sources as traced by the near-IR $K_s$-band emission (from 2MASS\footnote{https://www.ipac.caltech.edu/2mass/index.html}) according to a scale of $L_{0.5-2\,{\rm keV}}/L_{K_{\rm s}}=(4.7\pm0.4)\times10^{27}\,{\rm erg\,s^{-1}}\,L_{\odot}^{-1}$ \citep{ge15}.
Figure~\ref{fig:chandra} shows two intensity maps of hot plasma in two specific bands: 0.3--0.7 keV ($\sim$18--40 \AA) and 0.7--1.5 keV ($\sim$8--18 \AA).

The azimuthal-averaged surface brightness (SB) profiles of the two maps can be approximately characterized by the $\beta$-model \citep[e.g.][]{cavaliere76}: SB$(r)\propto(1+(r/r_c)^2)^{-3\beta+0.5}$, where $\beta$=0.46 and $r_c=1'.09$ for the 0.3--0.7 keV band, and $\beta$=0.55 and $r_c=1'.16$ for the 0.7--1.5 keV band, respectively (Figure~\ref{fig:bulge}a).
The X-ray emission in the 0.3--0.7 keV band has a more widespread distribution.
Under the spherical and isothermal assumptions, the density profile can be inferred as: $n(r)=n_0(1+(r/r_c)^2)^{-3\beta/2}$.
The inference of the central density $n_0$ depends on the emissivity function and hence on the plasma temperature \citep[e.g.][]{ge16}.
We find $n_0$=0.10 (or 0.078) cm$^{-3}$ for a temperature of 0.2 (or 0.3) keV, appropriate for the 0.3--0.7 keV (or 0.7--1.5 keV) band.
For comparison, Figure~\ref{fig:bulge}a shows the SB profile of the $K_s$-band emission along the major axis of M31, measured from a strip region with a narrow width of $1'$.
The $K_s$-band profile represents the already subtracted contribution of the unresolved X-ray sources, which peaks within the central $1'$ while the hot gas profile is more flattened.

\subsection{RGS Spectra}
The RGS observations and data reduction have been detailed in Section 4 of \citet{chen18}.
Briefly, there are 36 \xmm~observations toward the bulge region of M31, with a total effective exposure of 766 ks, as listed in Table~\ref{tab:log} in Appendix A. 
The first-order RGS spectra are extracted through standard pipeline {\tt ``rgsproc''}, taking the central coordinates of M31 (R.A.: 00$\rm ^h$42$\rm ^m$44$\rm ^s$.3503, decl.: +41$^{\circ}$16$'$08$''$.634, J2000; from NED) as a dispersion reference position.
The telescope pointing positions are all within a radius of $0'.5$ of the M31 center.
The corresponding background spectra are generated simultaneously from the blank-sky events. 
All the RGS spectra from individual observations are combined with the {\tt ``rgscombine''} script.

The RGS observation contains 1D spatial information in the cross-dispersion direction with a total width of $\sim5'$, where the FWHM of the line-spread function is about $0'.36$ around 20 \AA.
We extract a spectrum from a central ``broad region'' with a width of 4$'$ (920 pc) in each observation to maximize the signal to noise ratio.
Besides, we also extract spectra from three separate regions with different distances to the M31 center to study spatial dependent properties.
For each observation, the M31 center has been used as the reference point in the dispersion direction, and we only need to check where it is located in the cross-dispersion direction.
It can be derived by comparing the coordinates of the M31 center and the telescope pointing position that always anchors at the center in the cross-dispersion direction.
Then we consider the $\pm0'.3$ wide region around the M31 center as ``region a,'' the combination of the two $0'.7$ wide fields on both sides as ``region b,'' and the combination of two $1'$ wide fields farther out as ``region c'' (as illustrated in Figure~\ref{fig:chandra}).
The cross-dispersion directions of the 36 RGS observations are roughly along the minor axis of M31, with small variations within $\pm15^{\circ}$.
Under the spherical symmetric assumption of the hot gas, the spectra from different observations of these regions are combined separately.
The ``broad region'' spectrum is rebinned to have a bin size of 0.02 Å, while the spectra from ``region a, b, \& c'' have a bin size of 0.04 \AA.

Along the RGS dispersion axis, a spatial displacement translates into an apparent (first-order) wavelength shift of 0.138 \AA\ per arcmin in the first-order spectrum.
To account for this broadening effect, we convolve the angular structure function of a corresponding image with the model component when fitting the spectra, using the XSpec script {\tt rgsxsrc} (Figure~\ref{fig:bulge}b).
For example, the $K_s$-band image will be used for the unresolved point-source component, while the \chandra\ image will be used for the diffuse hot gas component, as detailed in Sec. 3.1.
Because the {\tt rgsxsrc} is designed for moderately extended ($\sim1'$) sources but the M31 bulge region is more extended, we modify these images a bit to make them more applicable.
We apply the telescope mirror vignetting effect to the images to account for the off-axis dimming retained in the RGS data.
And for the region where the RGS spectra are extracted, the same celestial region is cut from these images for generating reasonable angular structure functions.

\section{RGS spectral analysis}

\subsection{Fiducial fit with a single-T APEC}
\label{sec:fiducial}
We first characterize the diffuse hot plasma in each region with a single-temperature APEC model as a fiducial fit.
The contribution from the bright point sources, which are mostly low-mass X-ray binaries, is fitted with a featureless power-law model, while that from the unresolved point sources, mainly cataclysmic variables (CVs) and coronal active binaries (ABs), is represented by two {\it fixed} APEC components of the characteristic temperatures of 4.6 keV and 0.38 keV, respectively.
The luminosity of this unresolved-source emission is scaled to that of the stellar $K_s$-band emission in the same celestial region, and its RGS line broadening is also accounted by this $K_s$-band image via {\tt rgsxsrc}.
The foreground absorption with a column density of $6.7\times10^{20} \,\rm{cm^{-2}}$ \citep{dickey90} is applied to all the three radiating components.

\begin{figure*}[htbp] 
 \centering
        \includegraphics[angle=0,width=0.8\textwidth]{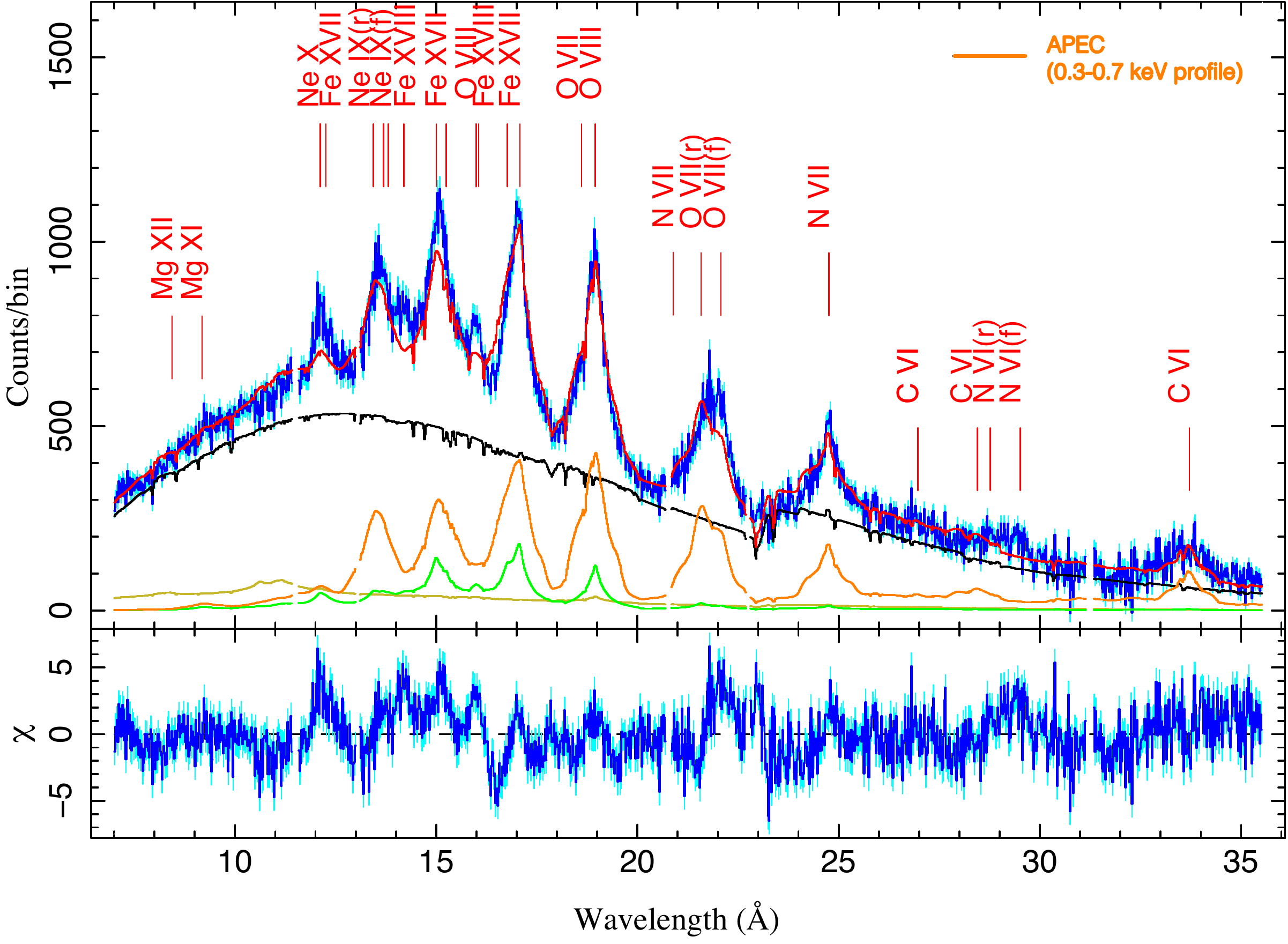}
        \includegraphics[angle=0,width=0.8\textwidth]{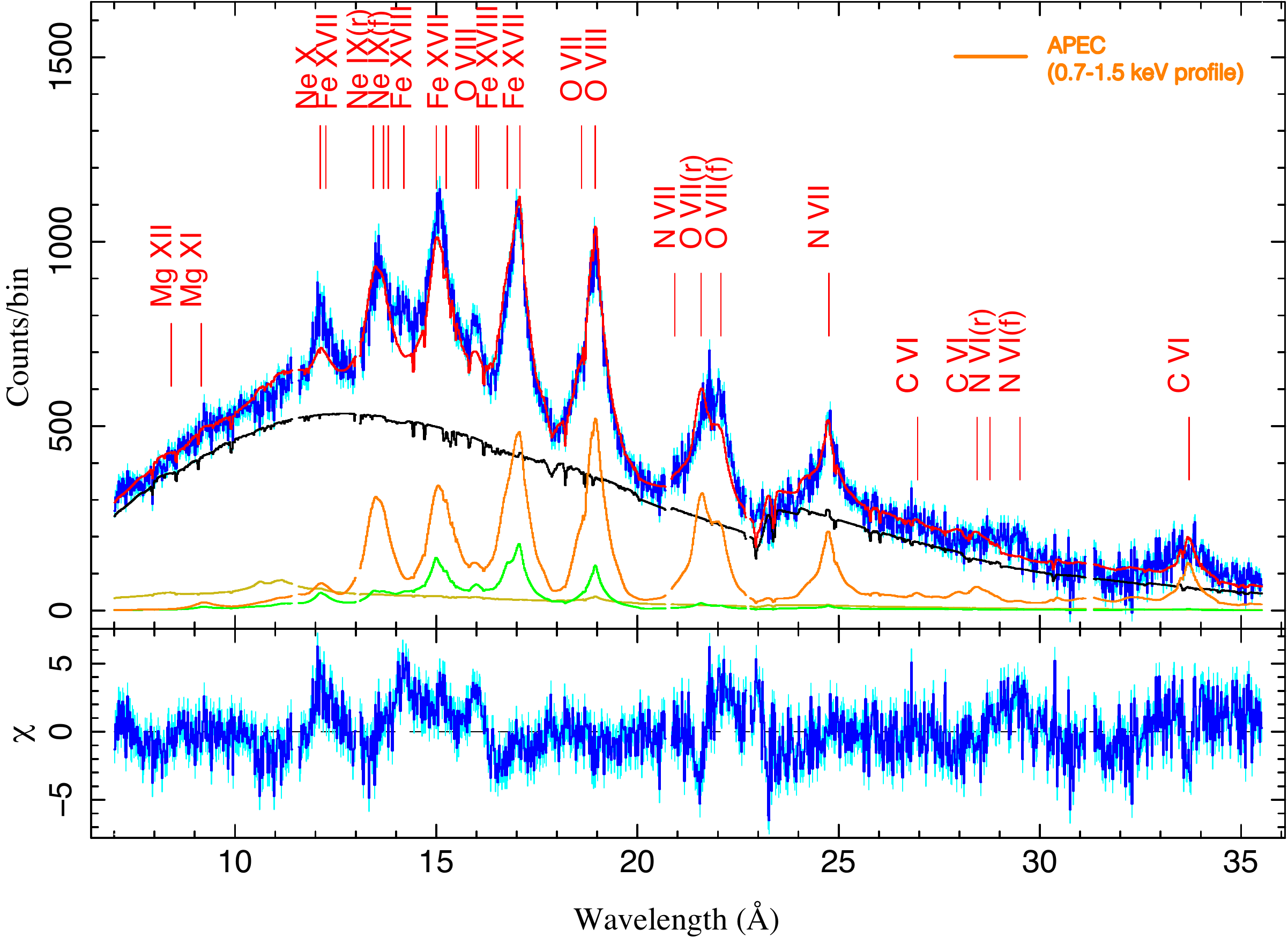}
            \caption{The ``broad region'' RGS count spectrum (the data are in blue, and their errors are in cyan) with a bin size of 0.02 \AA, and the fiducial fit (the best-fit model is in red) using a single-temperature APEC model for optically thin plasma that is convolved with the angular structure function of the 0.3--0.7 keV band (upper panel) or the 0.7--1.5 keV band (lower panel) \chandra\ image. The black curve represents the contribution from the bright point sources, the green and yellow curves show the contribution from the unresolved ABs and CVs, and the orange curve denotes the hot gas, respectively.}
\label{fig:apec} 
\end{figure*}

We use the \chandra\ images to account for the line broadening due to the spatial distribution of the hot plasma.
This broadening is energy-dependent, so the use of any image must be an approximation.
We try the images in the 0.3--0.7 and 0.7--1.5 keV bands, separately.
We find that the use of the 0.3--0.7 keV band image results in the line profile too broad to fit most emission lines of hot gas except for \cvi~Ly$\alpha$, as shown in Figure~\ref{fig:apec}.
In comparison, the profile from the 0.7--1.5 keV band image fits most lines well, and the $\chi^2$ of the fit decreases from 6276 to 5155 (Table~\ref{tab:pars}).
Therefore, we use this latter image as a first-order approximation.

The fitting results are listed in Table~\ref{tab:pars}.
The fitted temperatures are all a bit higher than 0.2 keV.
The temperatures in ``region b'' and ``region c'' are nearly the same, but is significantly smaller than that in ``region a.''
In all regions, the metallicity abundance of O is several times lower than that of other species such as C, N, Ne, and Fe, in the solar unit.

Not surprisingly, the above APEC modeling captures the bulk of the emission lines (e.g. Figure~\ref{fig:apec}), but the fits are by no means satisfactory.
There are significant deviations of the data from the model across the covered wavelength range (7-35.5 \AA).

First, we check the diagnostic He$\alpha$ triplets of the key He-like ions.
The strong \ovii~{\it f} line, which was previously identified in \citet{liu10}, is still prominent in the present RGS spectrum of an effective exposure eight times longer.
Covering a broader wavelength range than before, the RGS spectrum reveals a prominent forbidden line in the \nvi~triplet as well.
In addition, there is an excess around 21.8 \AA\ between the \ovii~{\it r} and {\it f} lines, and a small excess around 29 \AA\ between the \nvi~{\it r} and {\it f} lines.
The \neix~triplet also peaks between the {\it r} and {\it f} lines.

Next, we examine several other significant deviations from the APEC model.
(1) The RGS \nex~Ly$\alpha$ line is substantially stronger than the model line.
(2) The \fexviii~line at 14.2 \AA\ is not fitted at all.
(3) The \fexvii~lines around 15 \AA\ are stronger than the values predicted by the APEC model, and the observed peak of this line complex appears at the right side of the model peak.
(4) The \oviii~Ly$\beta$ line is much stronger than the model line, while the \cvi~Ly$\alpha$ seems broader than the APEC model line.

\begin{deluxetable*}{c|cc|ccc}
\tablecolumns{6}
\small
\tablewidth{0pt}
\tablecaption{The best-fit parameters of the fiducial fit}
\tablehead{ \colhead{} &  \multicolumn{2}{c}{Broad Region} &  \colhead{Region a} &  \colhead{Region b} &  \colhead{Region c}  }
\startdata
 & \multicolumn{5}{c}{Bright X-ray Binaries (Power-law)} \\
 \hline\noalign{\smallskip}
Norm & $4.58(\pm0.02)\times10^{-3}$&  $4.04(\pm0.02)\times10^{-3}$ & $1.10(\pm0.01)\times10^{-3}$ & $1.58(\pm0.01)\times10^{-3}$  & $1.33(\pm0.01)\times10^{-3}$   \\
$\Gamma$ &  $2.42\pm0.01$  &  $2.19\pm0.01$ & $1.88\pm0.02$ & $2.18\pm0.01$ & $2.39\pm0.02$  \\
\hline\noalign{\smallskip}
 & \multicolumn{5}{c}{ABs and CVs (fixed APECs)} \\
 \hline\noalign{\smallskip}
Norm$_{\rm (ABs)}$  &     $2.55\times10^{-4}$ &     $2.55\times10^{-4}$  &    $6.0\times10^{-5}$  &   $9.6\times10^{-5}$  &   $9.4\times10^{-5}$  \\
$kT_{\rm (ABs)}$    &     0.38 keV &     0.38 keV &0.38 keV &0.38 keV & 0.38 keV\\
Norm$_{\rm (CVs)}$    &   $1.25\times10^{-3}$ &   $1.25\times10^{-3}$ &  $3.0\times10^{-4}$  &  $4.7\times10^{-4}$  &  $4.6\times10^{-4}$  \\
$kT_{\rm (CVs)}$   &     4.6 keV &     4.6 keV &   4.6 keV &  4.6 keV  &  4.6 keV    \\
\hline\noalign{\smallskip}
& \multicolumn{5}{c}{Diffuse Hot Gas (APEC)} \\
Convolve with &  \tablenotemark{a}0.3--0.7 keV band & \tablenotemark{a}0.7--1.5 keV band & \multicolumn{3}{c}{0.7--1.5 keV band \chandra\ image} \\
 \hline\noalign{\smallskip}
Norm &  $1.44(\pm0.03)\times10^{-2}$ & $1.84(\pm0.03)\times10^{-3}$  &  $1.5(\pm0.1)\times10^{-4}$  &  $4.2(\pm0.1)\times10^{-4}$   &  $7.2(\pm0.2)\times10^{-4}$   \\
$kT$ [keV]   &    $0.201\pm0.002$    &   $0.226\pm0.001$   &   $0.246\pm0.001$  &   $0.223\pm0.004$   &   $0.224\pm0.003$  \\
C      &    $1.02\pm0.06$  	& $2.6\pm0.2$ 	& $4.5\pm0.8$  & $3.8\pm0.4$  & $3.5\pm0.3$  \\
N 	&    $0.28\pm0.02$  & $1.4\pm0.1$	& $4.4\pm0.4$ & $2.4\pm0.2$  & $1.5\pm0.1$  \\
O 	&    $0.108\pm0.002$& $0.42\pm0.01$ & $1.2\pm0.1$ & $0.74\pm0.04$ & $0.39\pm0.03$   \\
Ne    &  $0.37\pm0.02$ & $1.5\pm0.1$ & $4.4\pm0.3$ & $2.5\pm0.1$ & $1.3\pm0.1$ \\
Fe    &    $0.57\pm0.02$ & $1.6\pm0.1$	& $3.5\pm0.3$ & $3.1\pm0.2$  & $1.4\pm0.2$ \\
 \hline\noalign{\smallskip}
 $\chi^2$/d.o.f.    &   6276/(1391-9)=4.54    & 5155/(1391-9)=3.73  & 1688/(692-9)=2.47  & 2370/(692-9)=3.47 & 2431/(692-9)=3.56   \\
\enddata
\tablenotetext{a}{The images in the 0.3--0.7 or 0.7--1.5 keV band are used for the line broadening of the APEC model, separately.}
\tablecomments{The ``norm'' parameters are Xspec defaults \citep{arnaud96}. For the power-law model, the ``norm'' has the physical meaning of $\rm photons\,keV^{-1}cm^{-2}s^{-1}$ at 1 keV. For the APEC models, the `norm' has the physical meaning of $\frac{10^{-14}}{4\pi[D_{\rm A}(1+z)]^2} \int n_{\rm e}n_{\rm H} dV$, where $D_{\rm A}$ is the angular diameter distance to the source (cm), $z$ is the redshift, $n_{\rm e}$ and $n_{\rm H}$ are the electron and H densities (cm$^{-3}$), and $V$ is the volume.}
\label{tab:pars}
\end{deluxetable*}

\newpage

\subsection{Line Measurements}
We improve the flux measurements for $\sim$30 individual lines (Ne, O, N, C, and Fe) to quantify the deviations from the APEC model and to allow for spectroscopic analysis. The contribution from the point sources and the hot gas continuum emission is fixed to the values from the above fiducial fit.
Key lines are fitted simultaneously using narrow Gaussians ($\sigma\sim0.001$ \AA) convolving with the line broadening profiles derived from the 0.7--1.5 keV map as shown in Figure~\ref{fig:bulge}.
For the line with the double transitions, whose wavelength separation is far less than 0.01 \AA, we use only one single Gaussian.

We present in Appendix B a detailed description of lines included in the fit.
These lines are mainly Lyman series lines from H-like ions (\nex~Ly$\alpha$--Ly$\beta$, \oviii~Ly$\alpha$--Ly$\gamma$, \nvii~Ly$\alpha$--Ly$\delta$, and \cvi~Ly$\alpha$--Ly$\zeta$), K-shell lines from He-like ions (\neix~He$\alpha$, \ovii~He$\alpha$--He$\beta$, and \nvi~He$\alpha$), and iron lines (\fexvii~and \fexviii).
The measured line fluxes are listed in Table~\ref{tab:lines}, and the best-fit models are shown in Figure~\ref{fig:gauss}.

\subsubsection{Line Variations among Regions}
Regional variation of the line fluxes is examined from the inner to the outer regions in the M31 bulge.
According to the $\beta$-model derived from the 0.7--1.5 keV image, the relative fluxes of ``region a, b, and c'' should be 1:1.89:1.91 for an isothermal plasma.
For ease of comparison, Table~\ref{tab:lines} lists only half of the line fluxes measured in ``region b \& c.''
If the isothermal assumption is correct, the values should be about equal for the three regions, which is apparently not the case.

Different lines show divergent trends of the regional flux variation.
Some lines that require higher ionization have their fluxes decreasing from ``region a'' to ``c''; examples are the Lyman series lines of \nex~and \oviii.
The fluxes of both the \fexviii~(14.028 \AA) and the complex of \fexvii~lines around 17 \AA~are decreasing too, but the decreasing rate of the former is higher.
However, some lines that require lower ionization have the fluxes increasing from ``region a'' to ``c''; examples are the \cvi~Lyman series and the \nvi~He$\alpha$, as can be visually seen in Figure~\ref{fig:gauss}b.
Other lines, such as the K-shell lines of \neix~and \ovii, or the \nvii~Lyman series, which require somewhat medium ionization, do not show a general trend.

\subsubsection{Line Ratios}
Line ratios calculated based on the individual line fluxes are listed in Table~\ref{tab:ratios}.
We compare them with the theoretical CIE line-ratio curves in the temperature range of $10^6$ K to $2\times10^7$ K.

For H-like ions, the line ratios in each set of Lyman series show substantial inconsistency.
First, the Ly$\beta$/Ly$\alpha$ and Ly$\gamma$/Ly$\alpha$ ratios of \oviii\ are several times higher than the CIE predictions in all regions (Figure~\ref{fig:lyman}a), though the discrepancies decrease from ``region a'' to ``c.''
Second, the Ly$\beta$/Ly$\alpha$ of \nvii\ is a couple of times higher than the CIE prediction in ``region b'' or in ``broad region.''
Third, although the Lyman line ratios of \cvi\ are not well determined in ``region a'' and ``b,'' the Ly$\zeta$/Ly$\alpha$ in ``region c'' or ``broad region'' is nearly the same as the Ly$\gamma$/Ly$\alpha$ , which is not expected in a CIE case.

For He-like ions, we also find the G-ratios of \ovii, \neix, and \nvi~triplets all higher than the CIE predictions in nearly all regions (Figure~\ref{fig:cie}), except for the \neix\ G-ratio in ``region a'' and the \nvi\ G-ratio in ``region c''.

The line ratios of He- to H-like ions of one species are useful tracers for the plasma temperature. 
As shown in the middle column of Figure~\ref{fig:cie}, the He$\alpha$/Ly$\alpha$ ratios of O and N suggest  a temperature around $2\times10^6$ K, which decreases from ``region a'' to ``c''. 
The same ratio of Ne in ``broad region'' indicates a slightly higher temperature around $3\times10^6$ K.
The \nvii/\oviii~ratio, however, suggests a temperature less than $2\times10^6$ K, unless the abundance of nitrogen is several times higher than that of oxygen.

The inconsistency continues with the iron line ratios.
The immediate detection of the \fexviii~line needs a temperature higher than $\sim 4\times10^6$ K. 
The \fexviii~(14.208 \AA)/\fexvii~(17 \AA) ratios, which have small error bars, suggest a temperature of $\sim7\times10^6$ K, in all four regions.
At the same time, the \fexviii~16.004 \AA\ and 16.071 \AA\ lines that should have 80\% flux of the \fexviii~14.208 \AA\ line in a CIE case seem negligible.
Furthermore, both \fexvii~15.261 \AA/15.014 \AA\  and \fexvii~15 \AA/17 \AA\ ratios are considerably high (based on Table~\ref{tab:lines}), compared with the CIE ratio curves.
The \fexvii~15 \AA/17 \AA\ ratio may suggest a much higher temperature (e.g. $>2\times10^7$ K), but it should not be true, otherwise many highly ionized iron lines, such as \fexix\ or \fexx\ will appear.
It probably indicates that the contamination from \oviii~Ly$\gamma$ (15.176 \AA) is significant, although it causes new problems for the \oviii~Lyman series line ratios.
Therefore, the iron line ratios suggest that the plasma is not isothermal and most likely not in a CIE state.

\begin{figure*}[htbp] 
 \centering
        \includegraphics[angle=0,width=0.8\textwidth]{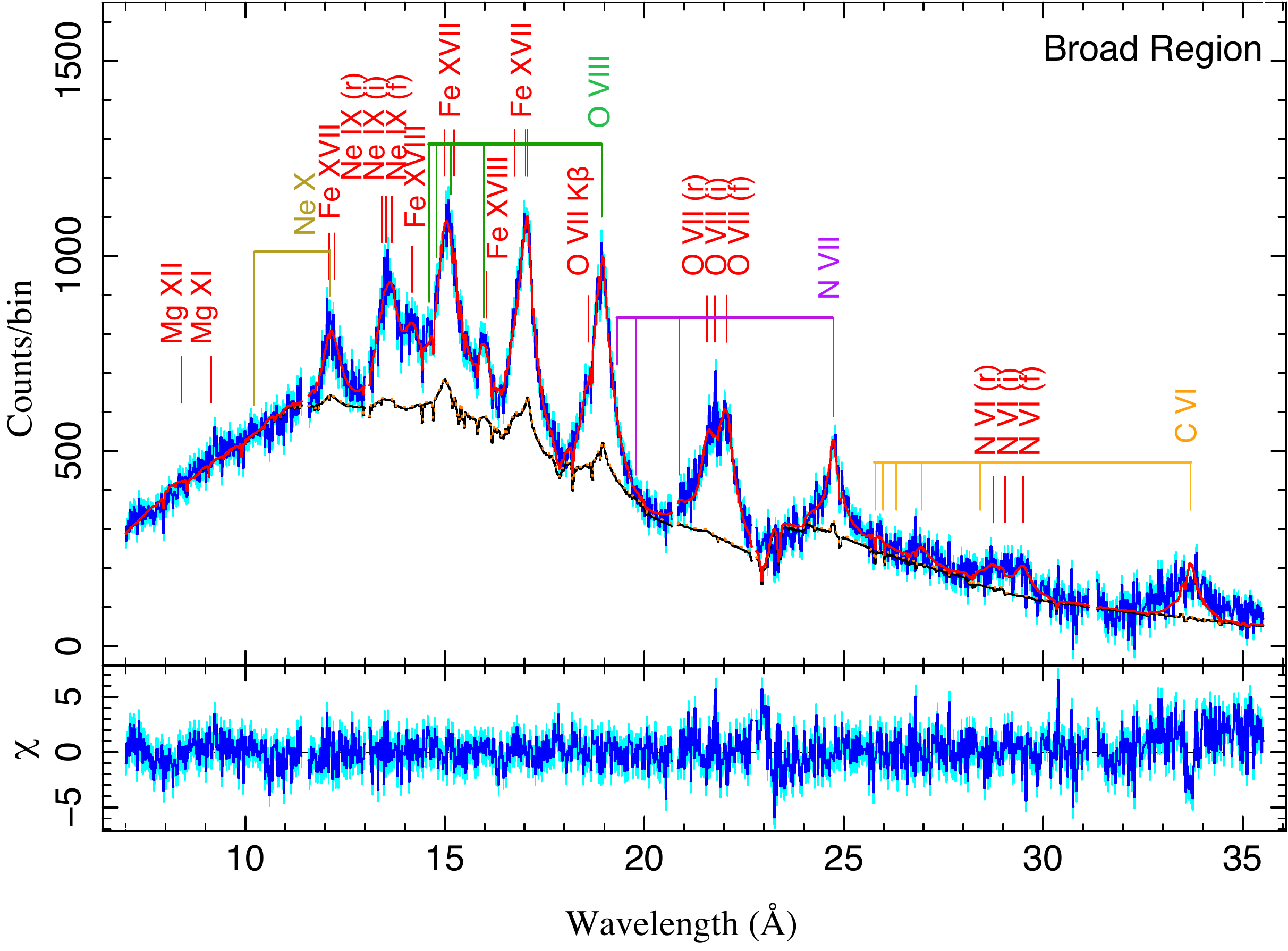}
\caption{(a) The same spectrum as in Figure~\ref{fig:apec}, but strong lines are fitted with multiple Gaussians.
The black curve represents the combined contribution from both bright and unresolved stellar sources determined in the fiducial fit, while the orange dotted curve adds the extra continuum emission from the APEC model, which is nearly negligible. The olive, green, purple, and yellow labels mark the Lyman series of \nex, \oviii, \nvii, and \cvi, respectively. (b) The spectra from ``region a, b, \& c'' with a bin size of 0.04 \AA.}
\label{fig:gauss} 
\end{figure*}

\setcounter{figure}{3}

\begin{figure*}[htbp] 
 \centering
        \includegraphics[angle=0,width=0.8\textwidth]{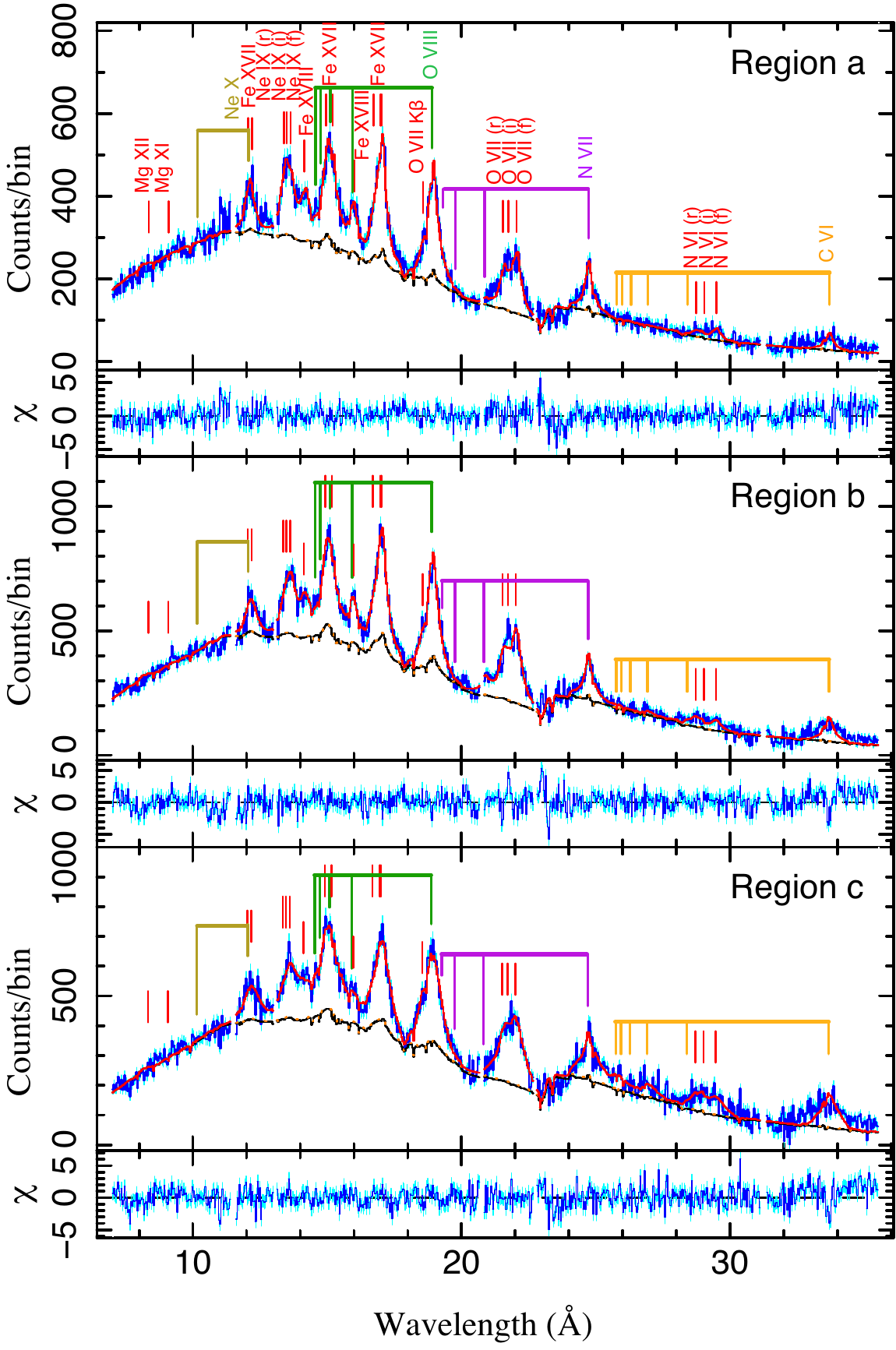}
            \caption{(Continued.)}
\label{fig:regions} 
\end{figure*}

\begin{deluxetable*}{lllr|cccc}
\tablecolumns{7}
\small
\tablewidth{0pt}
\tablecaption{Key line features}
\tablehead{\multirow{2}*{\bf Line } & \colhead{Rest $\lambda$}& \colhead{Oscillator} & \colhead{Level} &  \multicolumn{4}{c}{Flux ($10^{-5}\,{\rm ph\,s^{-1}\,cm^{-2}}$)}  \\
       & \colhead{(\AA)} & \colhead{Strength $f$}& \colhead{High--low}  &  \colhead{Broad region} &  \colhead{Region a} &  \colhead{Region b/2} &  \colhead{Region c/2}  }
\startdata
\multirow{2}*{\nex~Ly$\beta$} & 10.2385 & 5.267e$-$02 & 7--1  & \multirow{2}*{$<$0.4} & \multirow{2}*{$<$0.3} & \multirow{2}*{$<0.1$} & \multirow{2}*{$<$0.1}  \\
                           & 10.2396 & 2.624e$-$02 & 6--1  &  &  &   &  \\
\multirow{2}*{\nex~Ly$\alpha$} & 12.1321 & 0.276354& 4--1  & \multirow{2}*{1.9$\pm$0.5} & \multirow{2}*{1.8$\pm$0.2} & \multirow{2}*{$<0.1$} & \multirow{2}*{$<$0.4}  \\
                           & 12.1375 & 0.138116 & 3--1  &  &  &   &   \\
 \hline\noalign{\smallskip}
\neix~{\it r}   & 13.4473     & 0.741377   & 7--1   & 4.3$\pm$0.6   & 2.0$\pm$0.3  & 0.7$\pm$0.2 & 0.6$\pm$0.2  \\
\neix~{\it i}   & 13.5531     & 3.288e$-$4    & 5--1   & 1.9 ({\it f}/3.5)        & 0.4 ({\it f}/3.5)      &  0.4 ({\it f}/3.5)     & 0.4 ({\it f}/3.5)  \\
\neix~{\it f}   & 13.6990     & 8.246e$-$10  & 2--1   & 6.6$\pm$0.6   & 1.2$\pm$0.2  & 1.4$\pm$0.1 & 1.2$\pm$0.1  \\
 \hline\noalign{\smallskip}
 \hline\noalign{\smallskip}
\multirow{2}*{\fexvii}     & 12.124    & 0.460133   & 71--1  & 2.3$\pm$0.3         & $<$0.3                      & 0.6$\pm$0.1 & 0.7$\pm$0.1  \\
                                     & 12.266     & 0.393156  & 59--1  & 2.1 (above*0.92)  & $<$0.3 (above*0.92) & 0.5 (above*0.92)  & 0.6 (above*0.92)  \\
 \hline\noalign{\smallskip}
\multirow{2}*{\fexviii}   & 14.208      & 0.935159   & 56--1  & \multirow{2}*{6.1$\pm$0.6}   & \multirow{2}*{1.7$\pm$0.2}  & \multirow{2}*{1.2$\pm$0.2} & \multirow{2}*{0.9$\pm$0.2}   \\
             & 14.208                & 0.61436 & 55--1  &    &  &  &   \\
 \hline\noalign{\smallskip}
\multirow{2}*{\fexvii\tablenotemark{a}}    & 15.014                & 2.49408 & 27--1 & 11.4$\pm$0.7   & 2.7$\pm$0.3  & 2.4$\pm$0.2 & 2.7$\pm$0.2   \\    
             & 15.261                & 0.637919& 23--1  & 9.4$\pm$1.0  &  2.7$\pm$0.3  & 1.8$\pm$0.2  & 1.2$\pm$0.2   \\
 \hline\noalign{\smallskip}
\multirow{3}*{\fexvii}    & 16.780                 & 0.10473  & 5--1  & 3.1$\pm$0.8   & 1.2$\pm$0.3  & 0.6$\pm$0.2 & 0.5$\pm$0.2  \\
                                    & 17.051                & 0.12527  & 3--1  & 15.2$\pm$2.5 & 3.5$\pm$0.7  & 3.0$\pm$0.3 & 3.4$\pm$0.2  \\
                                    & 17.096                & 5.235e$-$8 & 2--1  & 5.9$\pm$1.1   & 1.9$\pm$0.5  & 1.6$\pm$0.2 & 0.4$\pm$0.3  \\
 \hline\noalign{\smallskip}
 \hline\noalign{\smallskip}
\multirow{2}*{\oviii~Ly$\gamma$\tablenotemark{a}} & 15.1760   & 1.931e$-$02 & 12--1   & \multirow{2}*{5.6$\pm$1.0}   & \multirow{2}*{1.8$\pm$0.3} & \multirow{2}*{1.0$\pm$0.2} & \multirow{2}*{0.3$\pm$0.2}  \\
                          & 15.1765   & 9.628e$-$03  & 11--1   &    &  &  &  \\
\multirow{2}*{\oviii~Ly$\beta$} & 16.0055   & 0.0524558& 7--1   & \multirow{2}*{7.0$\pm$0.5}   & \multirow{2}*{2.0$\pm$0.3} & \multirow{2}*{1.7$\pm$0.2} & \multirow{2}*{1.2$\pm$0.2}   \\
                          & 16.0067   & 0.0525078& 6--1   &    &  &  &   \\
\multirow{2}*{\oviii~Ly$\alpha$} & 18.9671  & 0.2770621& 4--1  &  \multirow{2}*{26.15$\pm$1.0}   &  \multirow{2}*{6.2$\pm$0.3} & \multirow{2}*{5.3$\pm$0.3} & \multirow{2}*{4.8$\pm$0.4}   \\
                           & 18.9725  & 0.1385085& 3--1  &     &  &  &  \\
  \hline\noalign{\smallskip}
\ovii~He$\beta$            &18.6270    & 0.1576123  & 13--1  & 4.2$\pm$0.7    & 0.9$\pm$0.3  &  0.8$\pm$0.2 & 1.3$\pm$0.2   \\
\ovii~{\it r}                         & 21.6015   & 0.7198880    & 7--1    & 13.1$\pm$0.9  & 2.4$\pm$0.4  & 2.5$\pm$0.3 & 2.9$\pm$0.3  \\     
\ovii~{\it i}                         & 21.8036      & 8.192e$-$05  & 5--1    & 5.7 ({\it f}/4.44)       & 1.1 ({\it f}/4.44)    & 1.2 ({\it f}/4.44) & 1.1 ({\it f}/4.44)   \\
\ovii~{\it f}                         & 22.0977      & 2.004e$-$10  & 2--1    & 25.2$\pm$0.9  & 5.1$\pm$0.3  & 5.5$\pm$0.2 & 4.5$\pm$0.3  \\
 \hline\noalign{\smallskip}
 \hline\noalign{\smallskip}
\multirow{2}*{\nvii~Ly$\delta$} & 19.3612     & 9.287e$-$03 & 19--1  &  \multirow{2}*{0.7$\pm$0.6}   & \multirow{2}*{0.5$\pm$0.2}  &\multirow{2}*{$<$0.4}  &  \multirow{2}*{$<$0.1} \\
                         & 19.3614     & 4.631e$-$03 & 18--1  &   &  &   &   \\
\multirow{2}*{\nvii~Ly$\gamma$} & 19.8257     & 1.932e$-$02 & 12--1  &  \multirow{2}*{$<$0.5}   & \multirow{2}*{$<$0.1}  &\multirow{2}*{$<$0.1}  &  \multirow{2}*{0.2}  \\
                         & 19.8261     & 9.636e$-$03 & 11--1  &   &  &   &   \\
\multirow{2}*{\nvii~Ly$\beta$} & 20.9095     & 5.269e$-$02 & 7--1  &  \multirow{2}*{3.3$\pm$0.8}   & \multirow{2}*{0.4$\pm$0.3}  &\multirow{2}*{1.1$\pm$0.2}  &  \multirow{2}*{$<$0.3}  \\
                         & 20.9106     & 2.630e$-$02 & 6--1  &   &  &   &  \\
\multirow{2}*{\nvii~Ly$\alpha$} & 24.7792 & 0.276944& 4--1  & \multirow{2}*{16.4$\pm$0.6}   & \multirow{2}*{3.6$\pm$0.3}   & \multirow{2}*{3.0$\pm$0.3} & \multirow{2}*{3.5$\pm$0.3}  \\
                           & 24.7846 & 0.138442& 3--1  &     &  &  &  \\
 \hline\noalign{\smallskip}
\nvi~{\it r} & 28.7870     & 0.7044293    & 7--1  & 9.1$\pm$1.6    & 1.1$\pm$0.5 & 1.4$\pm$0.4 & 2.8$\pm$0.4  \\
\nvi~{\it i} & 29.0843   & 3.523e$-$5     & 5--1  & 1.9 ({\it f}/6.85)       & 0.3 ({\it f}/6.85)    &  0.3 ({\it f}/6.85)  &  0.4 ({\it f}/6.85)   \\
\nvi~{\it f} & 29.5347   & 8.513e$-$11   & 2--1  & 12.7$\pm$1.5   & 2.2$\pm$0.6 & 2.0$\pm$0.4 & 2.8$\pm$0.5  \\
 \hline\noalign{\smallskip}
 \hline\noalign{\smallskip}
\multirow{2}*{\cvi~Ly$\zeta$} & 25.8302     & 3.208e$-$03 & 39--1       &  \multirow{2}*{3.1$\pm$0.9}   &  \multirow{2}*{$<$0.3} &  \multirow{2}*{$<$0.2}  & \multirow{2}*{1.5$\pm$0.3}  \\
                        & 25.8303      & 1.601e$-$03& 38--1   &     &  &  &  \\ 
\multirow{2}*{\cvi~Ly$\epsilon$} & 26.0260      & 5.197e$-$03 & 28--1   &  \multirow{2}*{$<$1.0}   &  \multirow{2}*{$<$0.4} &  \multirow{2}*{$<$0.2}  & \multirow{2}*{$<$0.2}  \\
                        & 26.0261      & 2.593e$-$03 & 27--1   &     &  &  &  \\ 
\multirow{2}*{\cvi~Ly$\delta$} & 26.3572      & 9.287e$-$03 & 19--1      &  \multirow{2}*{$<$0.7}   &  \multirow{2}*{$<$0.3} &  \multirow{2}*{$<$0.2}  & \multirow{2}*{$<$0.2}   \\
                        & 26.3574      & 4.635e$-$03 & 18--1   &     &  &  &  \\ 
\multirow{2}*{\cvi~Ly$\gamma$} & 26.9896  & 1.931e$-$02 & 12--1   &  \multirow{2}*{4.2$\pm$0.9}   &  \multirow{2}*{$<$0.4} &  \multirow{2}*{0.4$\pm$0.3}  &\multirow{2}*{1.6$\pm$0.3}  \\
                        & 26.9901     &  9.639e$-$03 & 11--1   &     &  &  &  \\ 
\multirow{2}*{\cvi~Ly$\beta$} & 28.4652      & 0.0525815 & 7--1       &  \multirow{2}*{2.4$\pm$1.3}   &  \multirow{2}*{$<$0.4} &  \multirow{2}*{$<$0.7}  & \multirow{2}*{$<$0.5}  \\
                        & 28.4663      & 0.0262897 & 6--1   &     &  &  &  \\ 
\multirow{2}*{\cvi~Ly$\alpha$} & 33.7342    & 0.277094& 4--1        & \multirow{2}*{48.0$\pm$2.7}   & \multirow{2}*{6.1$\pm$1.0}  & \multirow{2}*{7.8$\pm$0.8} & \multirow{2}*{12.3$\pm$0.8}  \\
                         & 33.7396    & 0.138525 & 3--1  &     &  &   &  \\
 \hline\noalign{\smallskip}
 \hline\noalign{\smallskip}
 \multicolumn{4}{c}{$\chi^2$/d.o.f.} &  2.63   &  1.84   &  2.33  &  2.63   \\
\enddata
\tablenotetext{a}{The fluxes of \oviii~Ly$\gamma$ are inferred from: \fexvii~[15.261 \AA]$-$\fexvii~[15.014 \AA]/3. Hereafter the \fexvii~[15 \AA] has this contribution excised.}
\tablecomments{According to the $\beta$-model derived from the 0.7--1.5 keV image, the relative fluxes of ``region-a,'' ``region-b/2,'' and ``region-c/2'' should be 1.0:0.945:0.955 for an isothermal CIE plasma.}
\label{tab:lines}
\end{deluxetable*}

\begin{deluxetable*}{lcccc}
\tablecolumns{5}
\small
\tablewidth{0pt}
\tablecaption{Key Line Ratios}
\tablehead{\colhead{Line Ratio} & \colhead{Broad Region}& \colhead{Region a} & \colhead{Region b} & \colhead{Region c}}
\startdata
\oviii~Ly$\beta$/Ly$\alpha$  & 0.268$\pm$0.022  &  0.323$\pm$0.051 &  0.321$\pm$0.042  & 0.250$\pm$0.047 \\
\oviii~Ly$\gamma$/Ly$\alpha$ & 0.214$\pm$0.039  &  0.290$\pm$0.050  &  0.189$\pm$0.039 & 0.063$\pm$0.042 \\
\nvii~Ly$\beta$/Ly$\alpha$ &  0.201$\pm$0.049  &  0.111$\pm$0.084  &  0.367$\pm$0.076   & $<$0.094  \\
\cvi~Ly$\beta$/Ly$\alpha$ &  0.050$\pm$0.027  &  $<$0.078  &  $<$0.100   &  $<$0.043 \\
\cvi~Ly$\gamma$/Ly$\alpha$ & 0.086$\pm$0.019  &  $<$0.078  &  0.051$\pm$0.039  & 0.130$\pm$0.026  \\
\cvi~Ly$\zeta$/Ly$\alpha$ & 0.065$\pm$0.019  &  $<$0.059  &  $<$0.029  &  0.122$\pm$0.026  \\
 \hline\noalign{\smallskip}
\ovii~G-ratio  &   2.36$\pm$0.18    &  2.58$\pm$0.46  &  2.68$\pm$0.34  &  1.93$\pm$0.24  \\
\neix~G-ratio &  1.98$\pm$0.36    &  0.80$\pm$0.18   &   2.57$\pm$0.76   &  2.67$\pm$0.91  \\
\nvi~G-ratio   &  1.60$\pm$0.34    &  2.27$\pm$1.21   &   1.64$\pm$0.57   &  1.14$\pm$0.26  \\
\ovii~He$\alpha$/\oviii~Ly$\alpha$  &  1.68$\pm$0.08  &   1.39$\pm$0.11   &  1.74$\pm$0.12  &  1.77$\pm$0.18 \\
\neix~He$\alpha$/\nex~Ly$\alpha$  & 6.74$\pm$1.85   &   2.00$\pm$0.31   &  $>$22.60  &  $>$4.91  \\
\nvi~He$\alpha$/\nvii~Ly$\alpha$    & 1.44$\pm$0.15    &   1.00$\pm$0.25  &  1.23$\pm$0.24  &  1.71$\pm$0.25   \\
\fexvii~(15 \AA/17 \AA) &  0.63$\pm$0.08  &   0.55$\pm$0.10  &   0.62$\pm$0.07  &  0.84$\pm$0.10  \\
\fexviii~(14.208 \AA)/\fexvii~(17 \AA) &  0.25$\pm$0.04  &  0.26$\pm$0.05  &  0.23$\pm$0.04  &  0.21$\pm$0.05  \\
\nvii~Ly$\alpha$/\oviii~Ly$\alpha$   &   0.63$\pm$0.03  &   0.58$\pm$0.06  &  0.57$\pm$0.07  &  0.73$\pm$0.09  \\
\enddata
\label{tab:ratios}
\end{deluxetable*}

\begin{figure}[htbp] 
\centering
\subfigure[]{
\begin{minipage}[b]{3.4in}
\centering
       \includegraphics[angle=0,width=\textwidth]{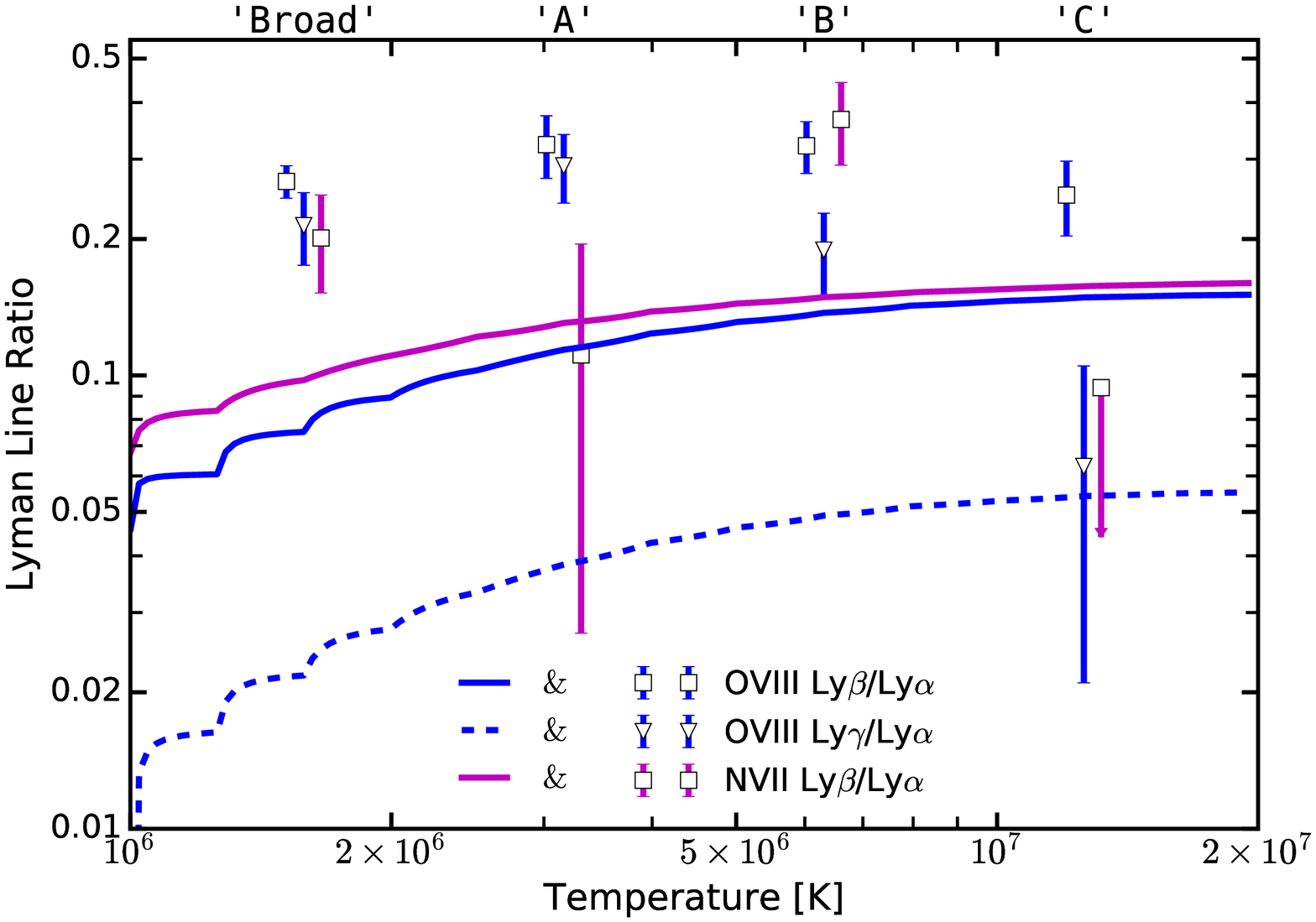} 
\end{minipage}%
}%

\subfigure[]{
\begin{minipage}[b]{3.4in}
\centering
       \includegraphics[angle=0,width=\textwidth]{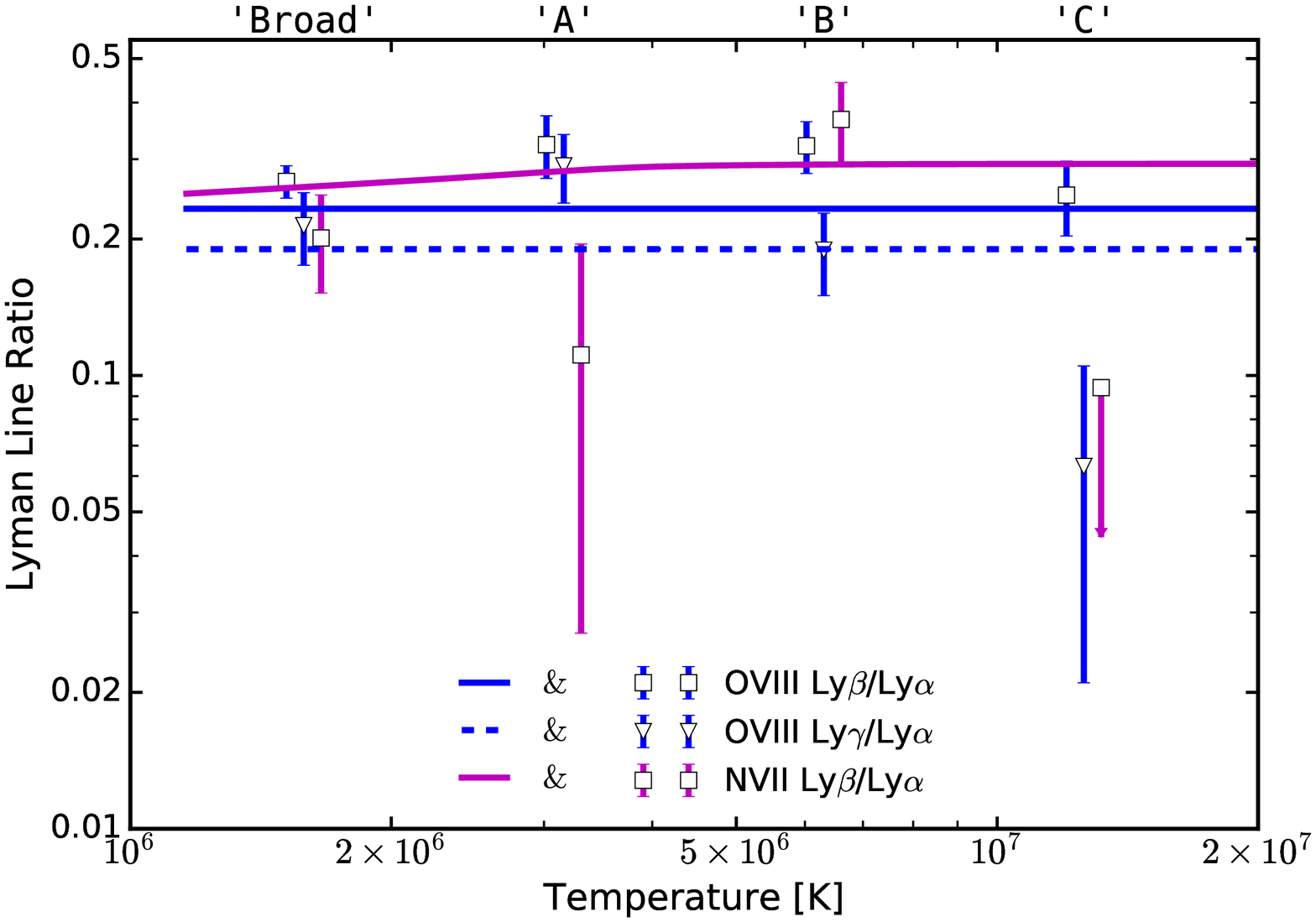} 
\end{minipage}%
}%

\subfigure[]{
\begin{minipage}[b]{3.4in}
\centering
       \includegraphics[angle=0,width=\textwidth]{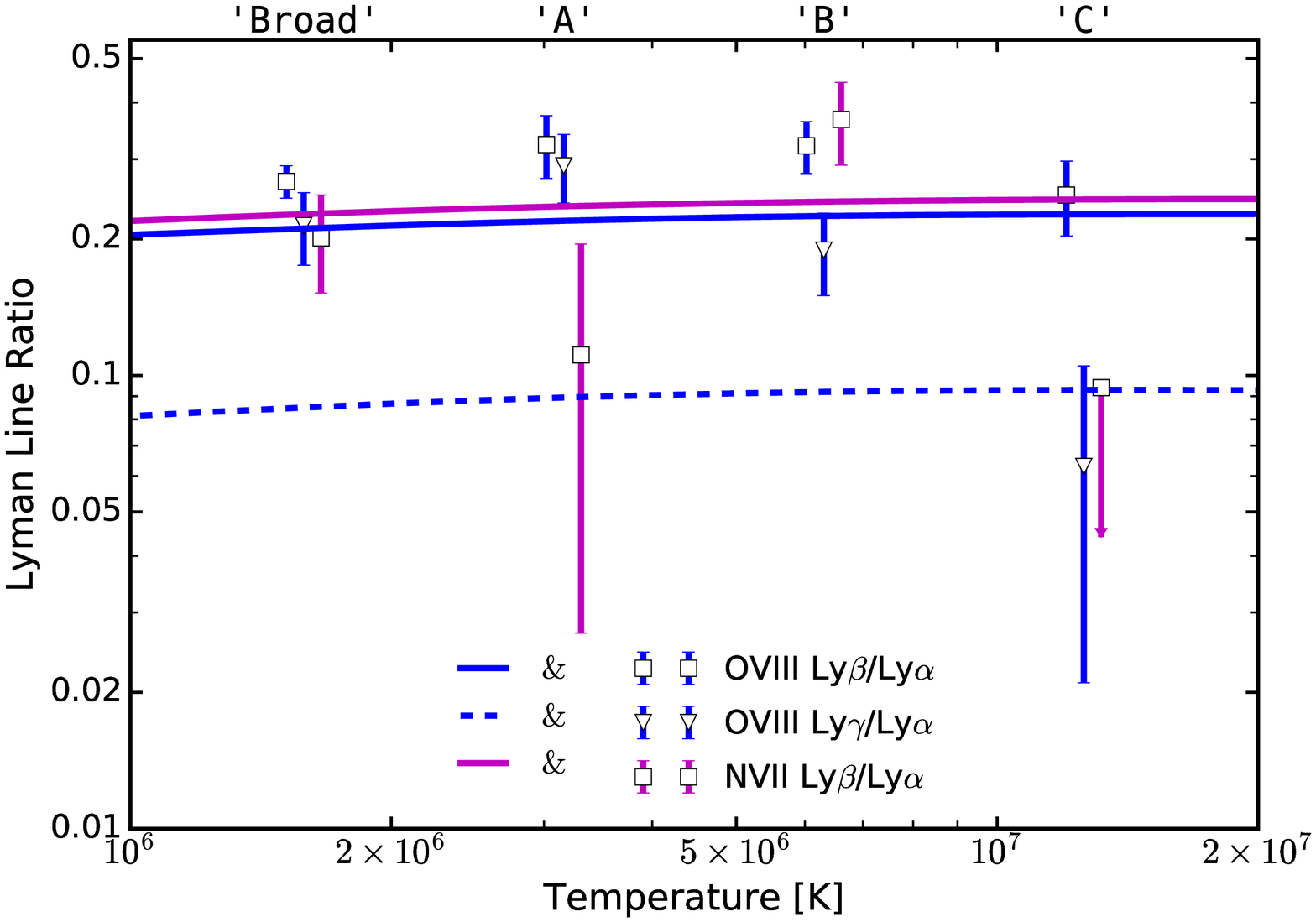} 
\end{minipage}%
}%
\caption{Lyman line ratios vs. plasma temperature. The four sets of data points are from the ``broad region'' and the ``region a to c,'' which are the same in each panel. The ratio curves, however, are calculated based on emission from a CIE plasma, from the CX process, and from the recombining process in the panel (a), (b), and (c), respectively.}
\label{fig:lyman} 
\end{figure}

\begin{figure*}[htbp] 
 \centering
        \includegraphics[angle=0,width=\textwidth]{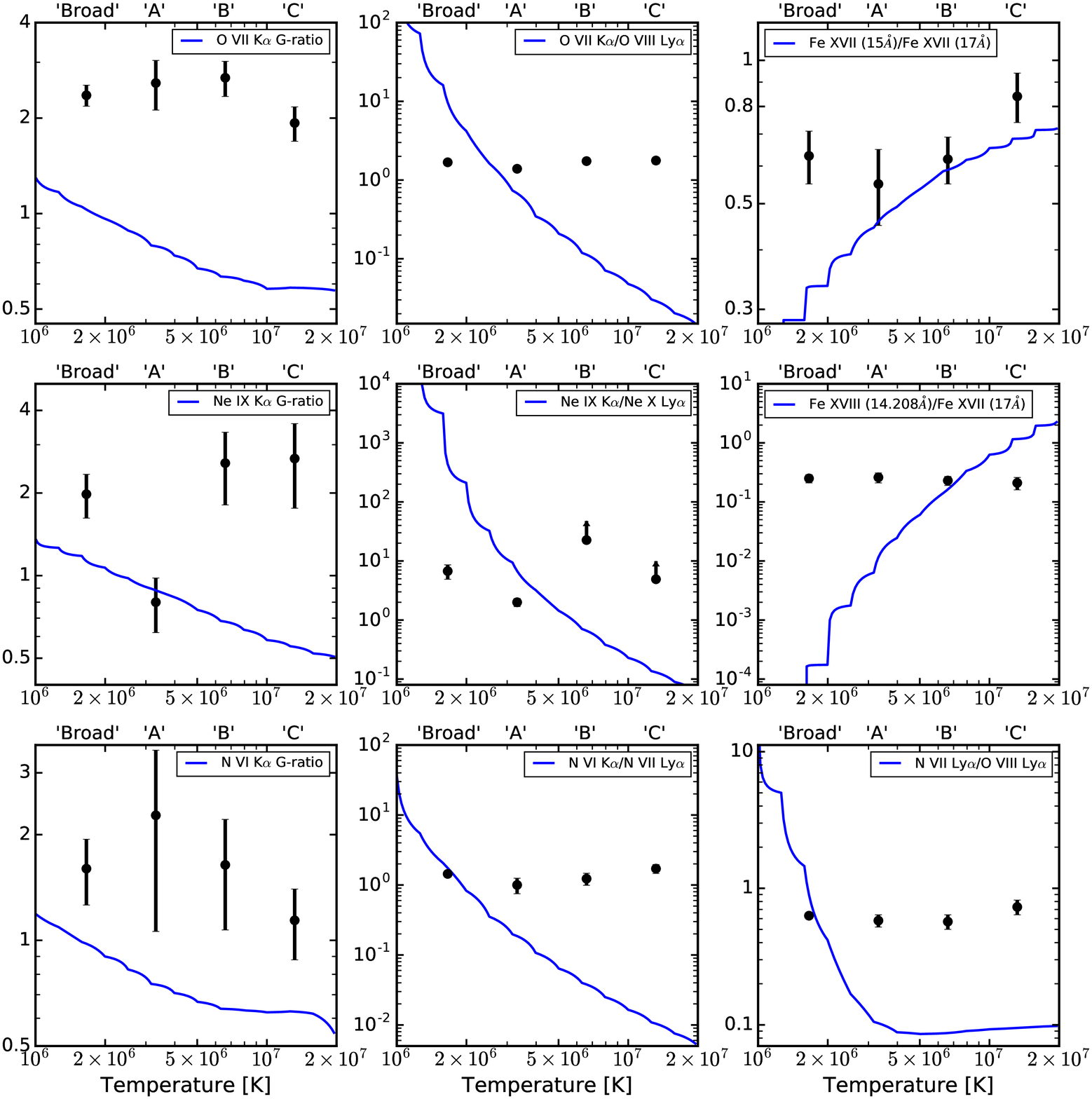} 
 \caption{Line ratios vs. plasma temperature. The four data points in each panel are from the ``broad region'' and the ``region a to c,''  respectively, while the blue curve is calculated for an optically thin thermal plasma under CIE.}
 \label{fig:cie} 
\end{figure*}

\subsubsection{Excess emission at 21.8 \AA}
The above spectral fit with multiple Gaussians still leaves a significant excess at 21.8 \AA.
This excess is apparent also in the coadded RGS1 spectrum without the application of {\tt ``rgscombine''}.
We check whether the excess is a physical or spurious feature.

We divide the 36 observations over 12 years (2001-2012; \citealt{chen18}) into four groups.
One group has the instrumental position angles of $\sim70^{\circ}$ and has a total effective exposure of 112 ks.
All other observations have the opposite position angles of $\sim250^{\circ}$.
We divide these latter observations into three groups according to the observing time so that they have comparable exposures of about 220 ks each.
The spectra in each group are combined by {\tt ``rgscombine''}.

\begin{figure*}[b] 
 \centering
        \includegraphics[angle=0,width=0.4\textwidth]{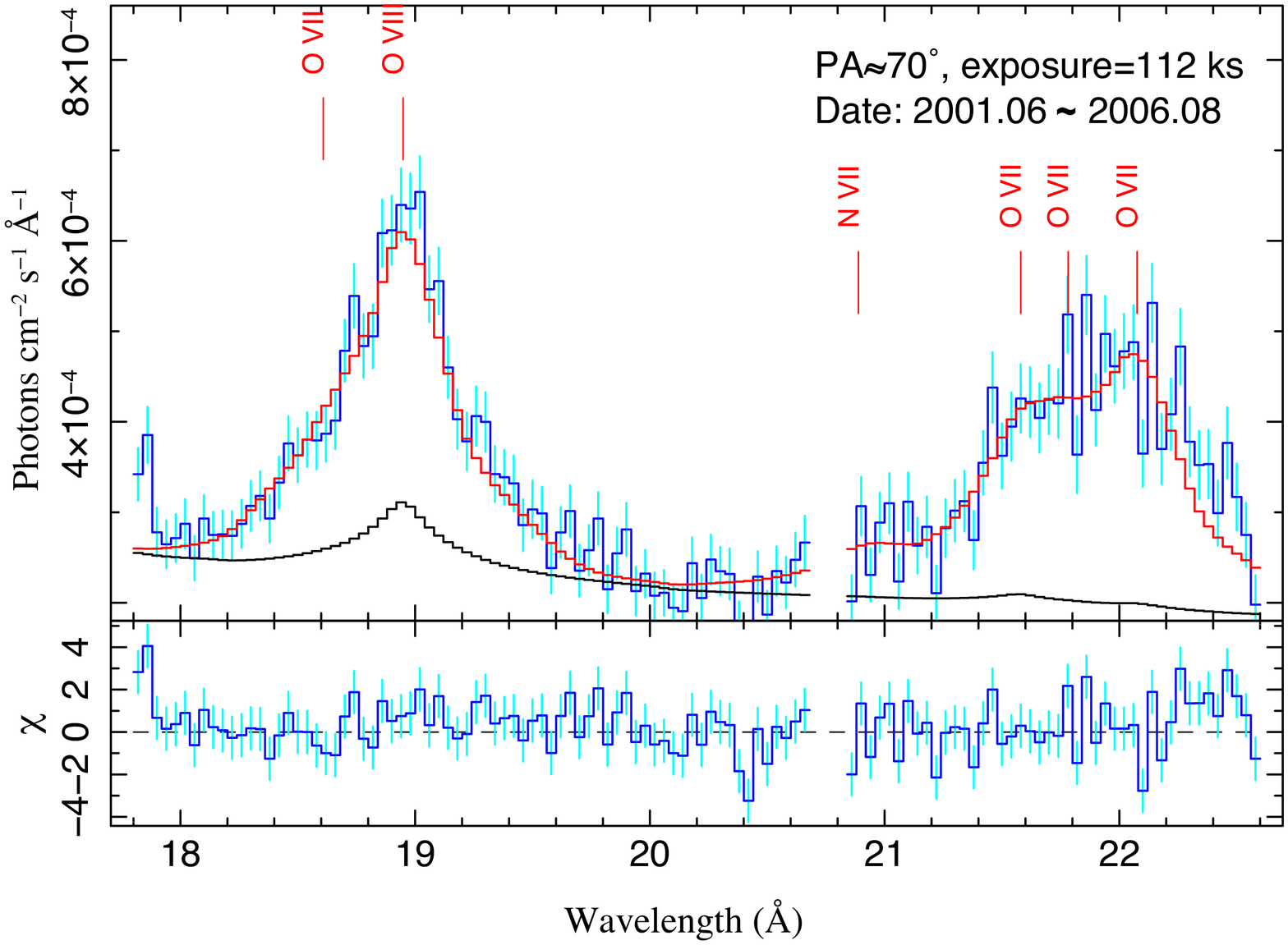}
        \includegraphics[angle=0,width=0.4\textwidth]{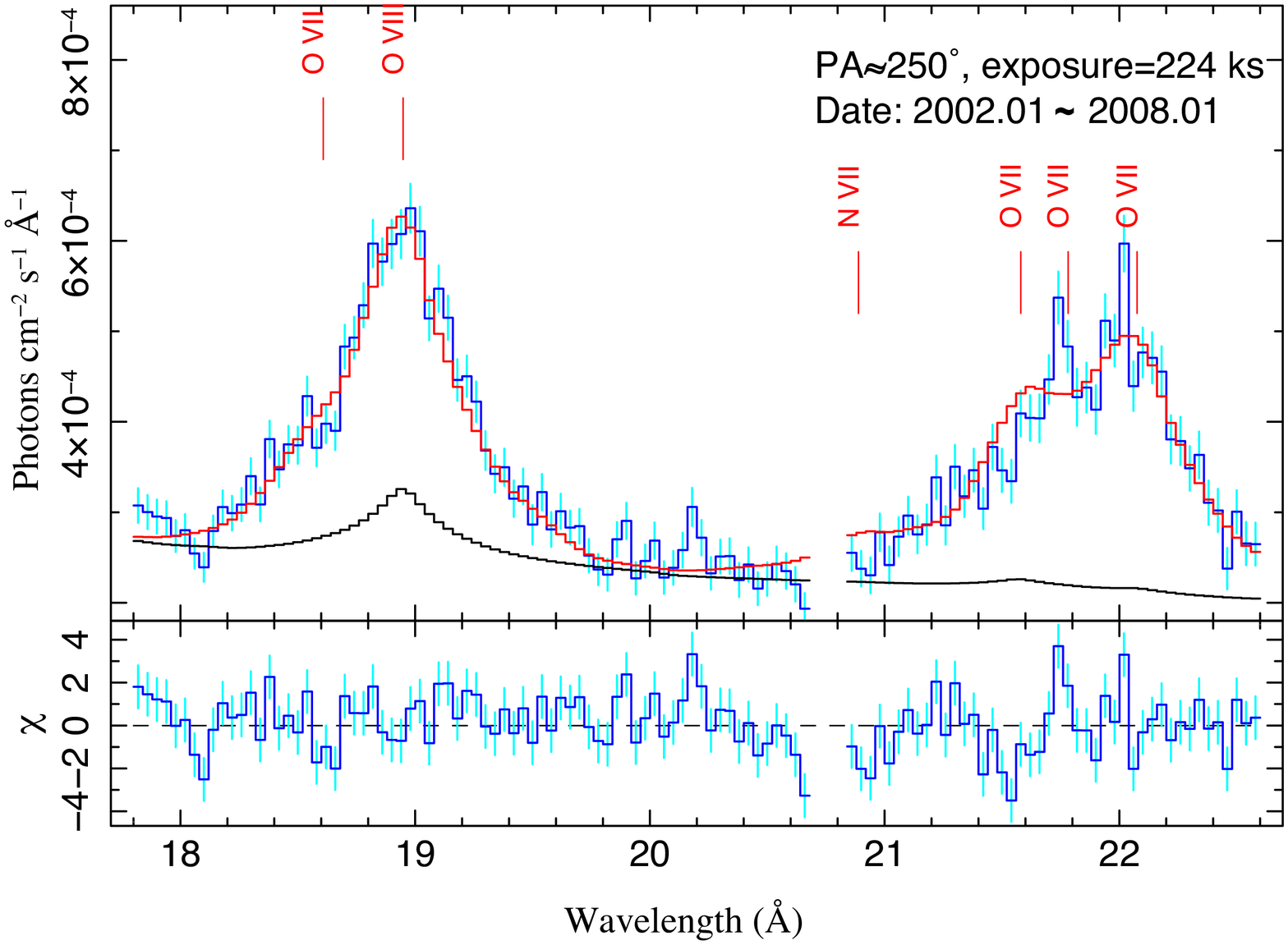}
        \includegraphics[angle=0,width=0.4\textwidth]{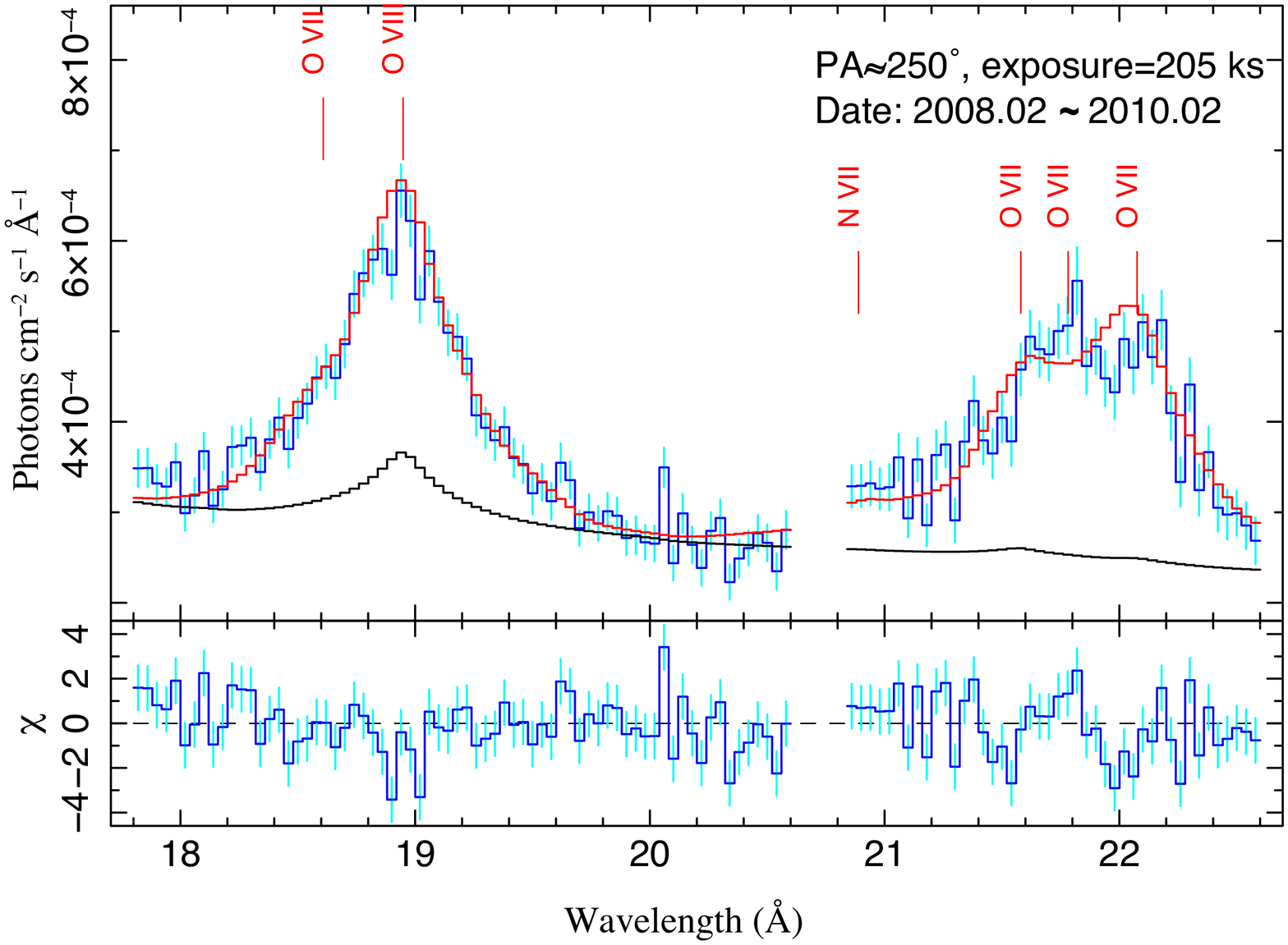}
        \includegraphics[angle=0,width=0.4\textwidth]{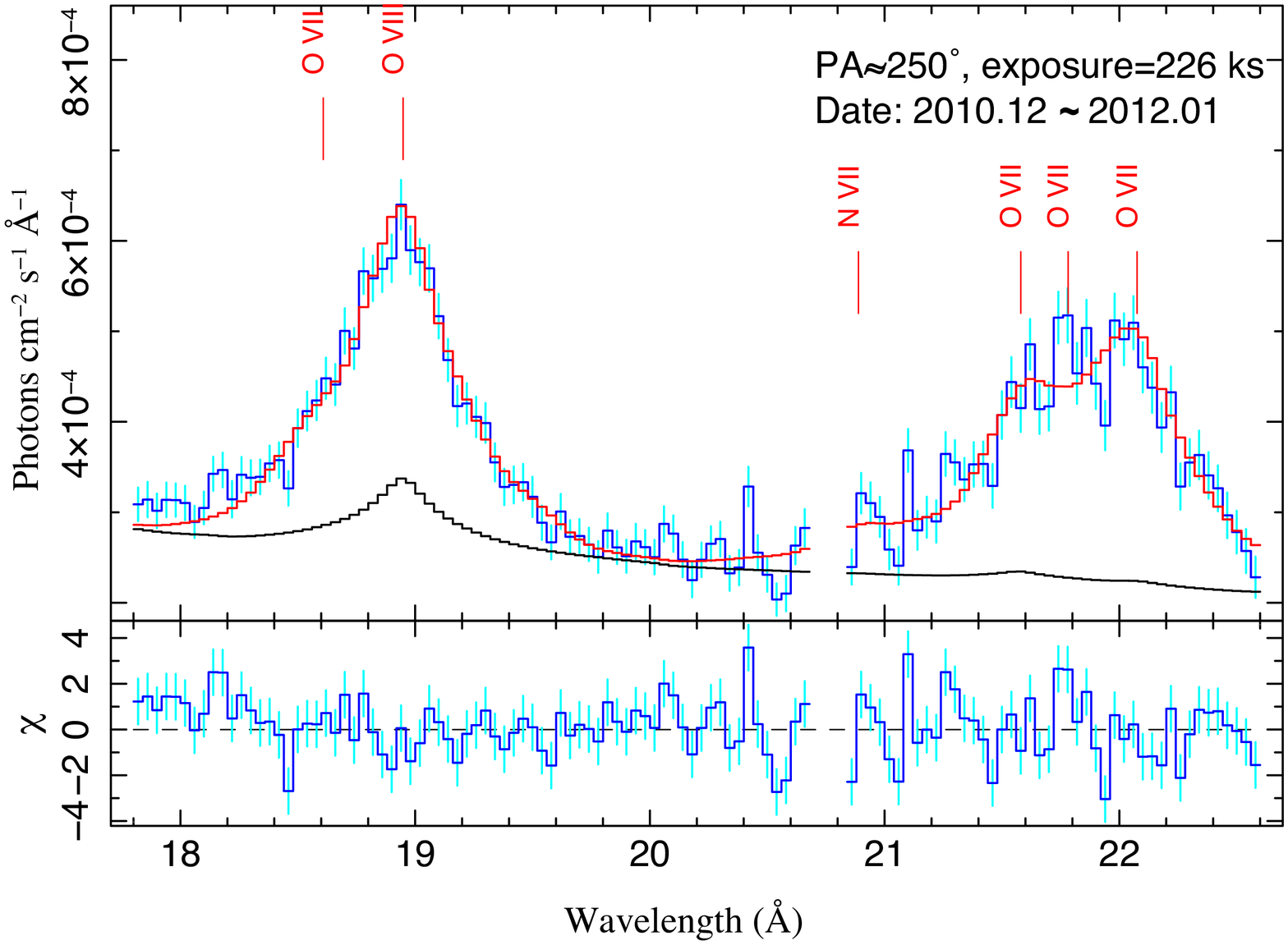}
            \caption{The \ovii/\oviii~complex and the 21.8 \AA\ excess. The upper-left panel shows the combined spectrum from the observations with the position angle of $\sim70^{\circ}$, while the other three panels show the spectra from observations with the position angle of $\sim250^{\circ}$, but in different epochs. The black curve represents the contribution of bright and unresolved point sources, and the red curve is the best-fit model, respectively.}
\label{fig:excess} 
\end{figure*}

Similarly, we do the fiducial fit with one-T APEC model first for the four spectra to obtain the continua of the bright point sources, whose brightness varies with time.
Once the emission from the point sources is obtained, the multiple Gaussians for lines as presented in Table~\ref{tab:lines} are superimposed onto that, unchanged because the time scale of the hot gas variation is supposed to be much longer than 12 yr.
The fitting results of the oxygen complex (18-23 \AA) of the four spectra are shown in Fig.~\ref{fig:excess}.

For the three spectra with position angles of $\sim250^{\circ}$, we calculate the Poisson probabilities of this 21.8 \AA\ excess.
In the first spectrum containing observation data from 2002 to 2008, the excess has a signal to noise ratio of $\sim$4.4 with the background taken into account.
The Poisson probability is 0.999995, and thus throughout this entire RGS spectrum there is a 0.15\% possibility to generate this feature by noises.
For the other two spectra observed during 2008-2010 and 2010-2012, the Poisson probabilities are $\sim$0.9998 and 0.99998, respectively.
So the feature is statistically significant.

\subsubsection{Non-isothermal and non-CIE signatures}
We summarize the significant non-isothermal and non-CIE spectral signatures in Table~\ref{tab:evaluation}.
These signatures are based on considerations from three aspects: line ratios, spatial variations of individual line intensities and line ratios, and shapes of line complexes.

Line ratios show both non-CIE and non-isothermal signatures.
For the non-CIE case, the G-ratios of He-like ions (He$\alpha$ triplets of \ovii, \nvi, and \neix) and the Lyman series line ratios (Ly$\beta$.../Ly$\alpha$ of \oviii~and \nvii) are higher than allowed values from a CIE plasma.
For the non-isothermal case, while the He- to H-like line ratios (He$\alpha$/Ly$\alpha$ of O, N, and Ne) would suggest a temperature of $\sim$0.2 keV, the iron line ratios (\fexviii/\fexvii, and \fexvii~(15 \AA/17 \AA)) indicate a much higher temperature of $\sim$0.7 keV, under the CIE assumption.

Both the line fluxes and the line ratios show clear spatial variations, which are the non-isothermal signs leastwise.
While the line fluxes of highly ionized species (e.g., \nex, \oviii, and \fexviii) tend to decrease from ``region a to c,'' the opposite trend applies to those lowly ionized ones (e.g., \cvi~and \nvi).
Both the Lyman series line ratios of \oviii~and the G-ratios of \ovii~and \nvi\ decrease from ``region a to c,'' while the He- to H-like line ratios of O and N increase from the inner to the outer region.
We should also bear in mind that the \nvii~Ly$\beta$/Ly$\alpha$ has the highest value in ``region b,'' and the \neix~G-ratio has a low value in ``region a.''

At last, the two excesses are the non-CIE features, while the broader \cvi~Ly$\alpha$ is the non-isothermal signature.
The 21.8 \AA\ excess is not expected in a CIE plasma because the \ovii~{\it i} line (21.80 \AA) usually is weak.
Though we have taken the \oviii~Ly$\gamma$ (15.176 \AA) as the reason for the excess around 15.2 \AA, this \oviii~line itself should not be that strong.

Any scenario introduced to explain the high \ovii~G-ratio in the M31 bulge should also account for all these non-isothermal and non-CIE signatures listed in the evaluation table (Table~\ref{tab:evaluation}).

\definecolor{orange}{rgb}{1,0.5,0.2}
\definecolor{olivegreen}{rgb}{0.24,0.5,0.19}

\begin{deluxetable*}{ll|l|l|l|l}
\tablecolumns{6}
\small
\tablewidth{0pt}
\tablecaption{Evaluation Form for Multiple Scenarios}
\tablehead{\multicolumn{2}{c}{non-CIE \& non-isothermal Features} & \colhead{Multi-T}& \colhead{CX} & \colhead{RS} & \colhead{Past-AGN}}
\startdata
\multirow{5}*{\rotatebox{90}{Line Ratio}}  & high \ovii~G-ratio & \textcolor{red}{$\times$:}  &  \textcolor{olivegreen}{$\surd$:} stronger {\it f} \& {\it i} line &  \textcolor{olivegreen}{$\surd$:} reduced {\it r} line  & \textcolor{olivegreen}{$\surd$:} stronger {\it f} \& {\it i} line \\
  & high \neix~G-ratio & \textcolor{red}{$\times$:}   &  \textcolor{olivegreen}{$\surd$:} stronger {\it f} \& {\it i} line  & \textcolor{orange}{$\circledcirc$:}  reduced {\it r} line & \textcolor{olivegreen}{$\surd$:} stronger {\it f} \& {\it i} line \\
  &high \nvi~G-ratio   & \textcolor{red}{$\times$:}   &  \textcolor{olivegreen}{$\surd$:} stronger {\it f} \& {\it i} line  &  \textcolor{red}{$\times$:} low opacity  & \textcolor{olivegreen}{$\surd$:} stronger {\it f} \& {\it i} line \\
  &high \oviii~Ly$\beta$/Ly$\alpha$ &  \textcolor{red}{$\times$:}  & \textcolor{olivegreen}{$\surd$:} enhanced  & \textcolor{olivegreen}{$\surd$:} reduced Ly$\alpha$  &  \textcolor{olivegreen}{$\surd$:} enhanced \\
  &high \oviii~Ly$\gamma$/Ly$\alpha$ & \textcolor{red}{$\times$:}  & \textcolor{olivegreen}{$\surd$:} enhanced &  \textcolor{olivegreen}{$\surd$:} reduced Ly$\alpha$ &  \textcolor{olivegreen}{$\surd$:} RRC and enhanced \\
  &high \cvi~Ly$\zeta$/Ly$\gamma$ & \textcolor{red}{$\times$:}  & \textcolor{olivegreen}{$\surd$:} enhanced & \textcolor{red}{$\times$:}  &  \textcolor{olivegreen}{$\surd$:} enhanced \\
  &high \nvii~Ly$\alpha$/\oviii~Ly$\alpha$ & \textcolor{orange}{$\circledcirc$:} low T &   \textcolor{red}{$\times$:} & \textcolor{olivegreen}{$\surd$:} reduced \oviii\ &  \textcolor{orange}{$\circledcirc$:} \\
  &high \fexvii~(15 \AA/17 \AA) & \textcolor{orange}{$\circledcirc$:} high T &  \textcolor{orange}{$\circledcirc$:} \fexvi~lines at 15 \AA  & \textcolor{red}{$\times$:} lower ratio  &   \textcolor{olivegreen}{$\surd$:} RRC and enhanced \\
  &high \fexviii/\fexvii~(17 \AA) & \textcolor{orange}{$\circledcirc$:} high T &  \textcolor{red}{$\times$:}    & \textcolor{red}{$\times$:}   &  \textcolor{olivegreen}{$\surd$:} RRC  \\
 \hline\noalign{\smallskip}
\multirow{4}*{\rotatebox{90}{Spatial Variation}}   &\nex~Lyman series decrease &  \textcolor{olivegreen}{$\surd$:} T decrease  &   \textcolor{orange}{$\circledcirc$:}  CX decrease &   \textcolor{red}{$\times$:}    &  \textcolor{olivegreen}{$\surd$:} less recombining \\
  &\oviii~Lyman series decrease &  \textcolor{olivegreen}{$\surd$:} T decrease   &   \textcolor{orange}{$\circledcirc$:}  CX decrease &   \textcolor{red}{$\times$:}    &  \textcolor{olivegreen}{$\surd$:} less recombining \\
  &\fexviii~(14.208 \AA) decrease &  \textcolor{orange}{$\circledcirc$:} T decrease  & \textcolor{red}{$\times$:}    &   \textcolor{red}{$\times$:}     &   \textcolor{olivegreen}{$\surd$:} less RRC \\
  &\fexvii~(17 \AA) decrease  &  \textcolor{olivegreen}{$\surd$:} T decrease & \textcolor{red}{$\times$:}   &   \textcolor{red}{$\times$:}     &  \textcolor{olivegreen}{$\surd$:} T decrease \\
  &\ovii~He$\alpha$--He$\beta$ no trend &  \textcolor{olivegreen}{$\surd$:} T decrease &  \textcolor{orange}{$\circledcirc$:}  CX stable  &  \textcolor{red}{$\times$:}    &   \textcolor{olivegreen}{$\surd$:} T decrease \\
  &\nvii~Lyman series no trend &  \textcolor{olivegreen}{$\surd$:} T decrease &  \textcolor{orange}{$\circledcirc$:}  CX stable & \textcolor{red}{$\times$:}    &   \textcolor{olivegreen}{$\surd$:} T decrease  \\
  &\nvi~He$\alpha$ increase &  \textcolor{olivegreen}{$\surd$:} T decrease  &  \textcolor{orange}{$\circledcirc$:}  CX increase  & \textcolor{olivegreen}{$\surd$:} scattered out  &  \textcolor{olivegreen}{$\surd$:} T decrease \\
  &\cvi~Lyman series increase &  \textcolor{olivegreen}{$\surd$:} T decrease  &  \textcolor{orange}{$\circledcirc$:}  CX increase & \textcolor{olivegreen}{$\surd$:} scattered out &  \textcolor{olivegreen}{$\surd$:} more recombining \\
 \cline{2-6}\noalign{\smallskip}
  &\oviii~Ly$\beta$/Ly$\alpha$ decrease & \textcolor{red}{$\times$:}   &   \textcolor{red}{$\times$:}     & \textcolor{olivegreen}{$\surd$:} Ly$\alpha$ out &  \textcolor{olivegreen}{$\surd$:} less recombining \\
  &\oviii~Ly$\gamma$/Ly$\alpha$ decrease & \textcolor{red}{$\times$:}   &   \textcolor{orange}{$\circledcirc$:}  CX decrease  & \textcolor{olivegreen}{$\surd$:} Ly$\alpha$ out &  \textcolor{olivegreen}{$\surd$:} less recombining \\
  &\nvii~Ly$\beta$/Ly$\alpha$ high in ``region b'' & \textcolor{red}{$\times$:}  &   \textcolor{red}{$\times$:}     &  \textcolor{red}{$\times$:}  &  \textcolor{olivegreen}{$\surd$:} enhanced \\
  &\neix~G-ratio low in ``region a'' & \textcolor{red}{$\times$:}   &  \textcolor{orange}{$\circledcirc$:}  CX not in ``region a'' &  \textcolor{red}{$\times$:}  &  \textcolor{olivegreen}{$\surd$:} spheral shell \& \fexix \\
  &\ovii~G-ratio decrease & \textcolor{red}{$\times$:}  &   \textcolor{orange}{$\circledcirc$:}  CX decrease & \textcolor{olivegreen}{$\surd$:} {\it r} line out &  \textcolor{olivegreen}{$\surd$:} less recombining \\
  &\ovii~He$\alpha$/\oviii~Ly$\alpha$ increase &  \textcolor{olivegreen}{$\surd$:} T decrease   &  \textcolor{red}{$\times$:}    &  \textcolor{orange}{$\circledcirc$:} {\it r} line out   &  \textcolor{olivegreen}{$\surd$:} T decrease \\
  &\nvi~He$\alpha$/\nvii~Ly$\alpha$ increase  &  \textcolor{olivegreen}{$\surd$:} T decrease  &  \textcolor{red}{$\times$:}    &  \textcolor{orange}{$\circledcirc$:} {\it r} line out  &  \textcolor{olivegreen}{$\surd$:} T decrease \\
  &\nvii~Ly$\alpha$/\oviii~Ly$\alpha$ increase &  \textcolor{olivegreen}{$\surd$:} T decrease  &  \textcolor{red}{$\times$:}    &  \textcolor{red}{$\times$:} inverse  &  \textcolor{olivegreen}{$\surd$:} T decrease  \\
 \hline\noalign{\smallskip}
\multirow{4}*{\rotatebox{90}{Shape}}   &\fexvii~(excess at 15.2 \AA)  & \textcolor{red}{$\times$:} peak 15.014 \AA & \textcolor{olivegreen}{$\surd$:} enhanced \oviii~Ly$\gamma$ & \textcolor{orange}{$\circledcirc$:} broader 15.014 \AA\ line & \textcolor{olivegreen}{$\surd$:} enhanced \oviii~Ly$\gamma$\\
  &\ovii~He$\alpha$ (excess at 21.8 \AA)    & \textcolor{red}{$\times$:} peak 21.60 \AA & \textcolor{red}{$\times$:}   & \textcolor{orange}{$\circledcirc$:} broader {\it r} line & \textcolor{olivegreen}{$\surd$:} spheral shell\\
  & broad \cvi~Ly$\alpha$ &  \textcolor{olivegreen}{$\surd$:} T decrease  &   \textcolor{red}{$\times$:}  & \textcolor{olivegreen}{$\surd$:} scattered out  &  \textcolor{olivegreen}{$\surd$:} T decrease \\
\enddata
\tablecomments{Major deviations from a single-temperature CIE plasma. \textcolor{olivegreen}{$\surd$} means that this signature can be explained by this scenario, \textcolor{orange}{$\circledcirc$} means that this scenario can alleviate the deviation, though not completely, and \textcolor{red}{$\times$} means that this scenario does not help to solve the deviation.}
\label{tab:evaluation}
\end{deluxetable*}

\section{CX and RS in a multiple-T plasma}
Here we examine the possibilities of the two very distinct scenarios proposed in previous works: CX \citep{liu10} and RS \citep{chen18}.
But first of all, there are lines of evidence, mostly from the spatial-variation results, suggesting that the plasma, if locally in a CIE state, has a decreasing temperature trend from the inner to the outer bulge region.
The allowance of this global temperature variation introduces more flexibility to the two scenarios to explain the spectroscopic signatures, compared with that in an isothermal plasma.

\subsection{CIE plasma with outward decreasing temperature}
The trend of the outward decreasing temperature is seen even in the fiducial fit (Table~\ref{tab:pars}).
Accounting for the projection effect, one should expect a more significant temperature difference from the inner to the outer region.
This scenario is also consistent with the measured variation of the diagnostic line fluxes.
The higher ionization lines decrease in flux from ``region a'' to ``c'' while the lower ionization lines increase.
This is because the emissivities of the former ion lines peak at higher temperatures than the latter.
What corresponds to the decreasing temperature most directly is that the line ratios of He$\alpha$/Ly$\alpha$ lines and \nvii~Ly$\alpha$/\oviii~Ly$\alpha$ show a slight but unambiguous increase from ``region a'' to ``c'' (Figure~\ref{fig:cie}).

The outward decreasing temperature scenario could also naturally explain why the 0.7--1.5 keV SB profile is steeper than the 0.3--0.7 keV one.
The reason is that the former band contains more higher ionization lines of Ne and Fe, which decrease in flux, while the latter band possesses more lower ionization lines of C, N, and O, whose fluxes increase from the inner to the outer region.

Of course, the scenario with the local CIE assumption cannot explain many features listed in Table~\ref{tab:evaluation}.
These features include the high G-ratios, the high ratios of Lyman series lines, and the 21.8 \AA\ excess.
Furthermore, the scenario is also not consistent with the high temperatures inferred by the iron lines.
The \fexviii/\fexvii~(17 \AA) ratio and \fexvii~15 \AA/17 \AA\ ratio (\oviii~Ly$\gamma$-corrected) suggest $\sim$0.7 keV and $\sim$0.4 keV in both the inner and outer regions.
Such high temperatures conflict with $T\sim0.2$ keV suggested by other line ratios in the same regions.
Therefore, a CIE plasma with a global temperature decrease with the distance from the galaxy's center is not sufficient. 
Additional supplementary physical processes need to be considered.

\subsection{Charge exchange (CX)}
\begin{figure*}[htbp] 
 \centering
        \includegraphics[angle=0,width=0.8\textwidth]{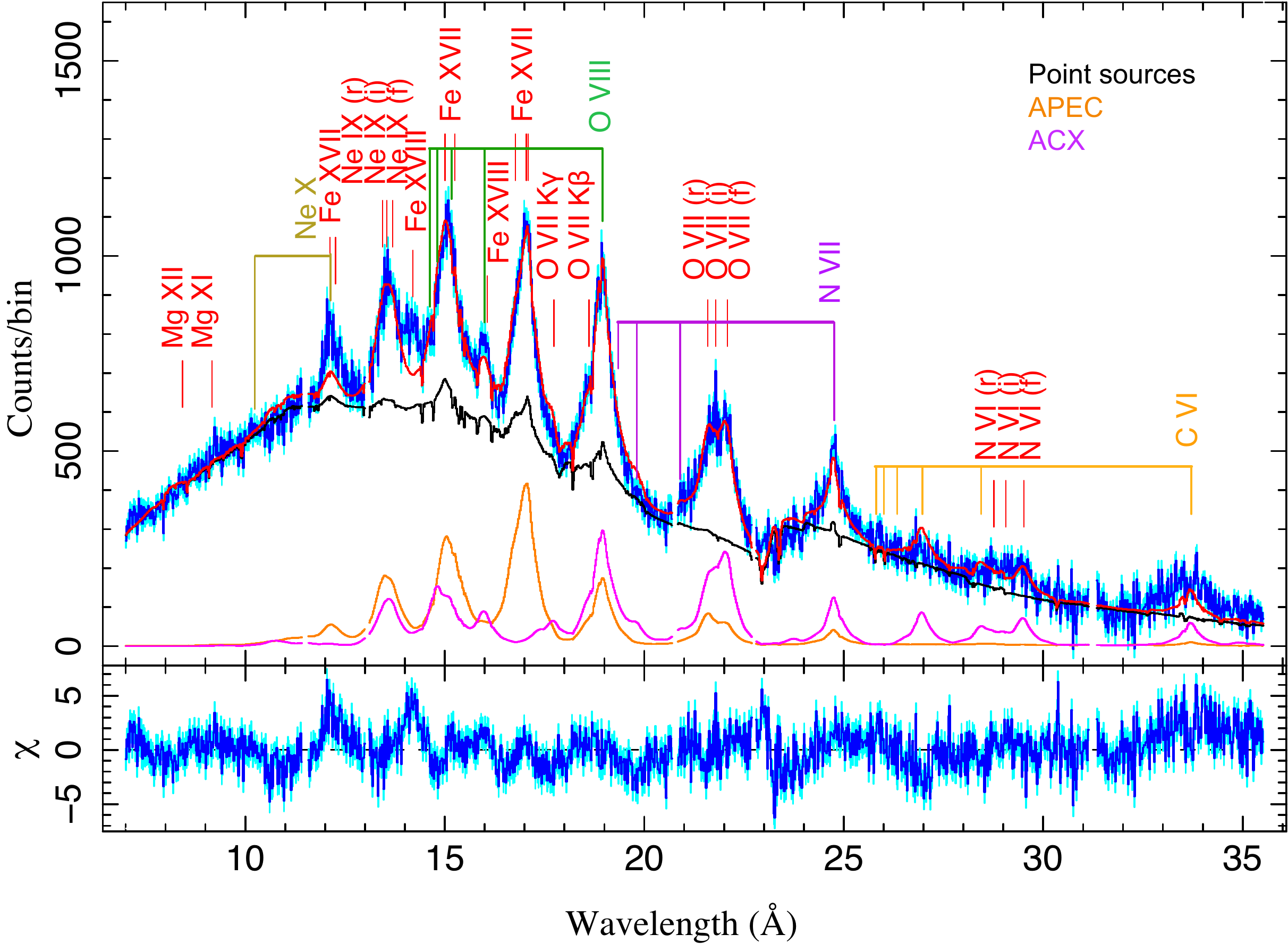} 
 \caption{The best fit of the ``broad region'' spectrum with an ACX model included. The components of point sources, APEC, and ACX are shown in black, orange, and magenta, respectively.}
 \label{fig:apecCX} 
\end{figure*}

CX may occur in the inner bulge of M31, e.g., at the interface between hot gas and the flattened spiral structure revealed by H$\alpha$ emission \citep[e.g.][]{dong16}.
We test this scenario by adding a CX component to the fiducial fit that has used a single-temperature APEC model to describe the hot gas (Sec. 3.1).
We use the ACX model v1.0.2\footnote{http://www.atomdb.org/CX/index.php} to account for the CX emission \citep{smith12}.
The inclusion of this model adds one free parameter ``norm,'' while the other ACX parameters, such as temperature, metal abundances, and redshift, are tied to those of the APEC model \citep[in a similar way as in][]{zhang14}.
Figure~\ref{fig:apecCX} presents the best-fit combined APEC+ACX model for the ``broad region'' spectrum.
The best-fit temperature and the oxygen abundance are $\sim$0.25 keV and $\sim$1.27 solar value respectively, higher than those from the fiducial fit.

CX enhances the G-ratios of the He-like He$\alpha$ triplets.
This enhancement is due to exchanged electrons populating the triplet states and eventually cascading into the 1s2s$\rm \sim ^3\!S_1$ level, which is the upper level of the {\it f} line.
As shown in Figure~\ref{fig:apecCX}, this APEC+ACX model fit the \ovii, \nvi, and \neix\ He$\alpha$ triplets well.
But the model does not increase the {\it i} line of the \ovii~triplet significantly, and cannot explain the 21.8 \AA\ excess.
The R-ratio ({\it f/i}) in the ACX model remains larger than 4 for a plasma with a temperature lower than 1 keV, which is similar to that of 4.44 in an APEC-only plasma.

Another spectroscopic diagnostic that cannot be fully explained by CX is the enhanced transitions from higher $n$ shells.
Indeed, both \oviii~Ly$\beta$ and Ly$\gamma$ and \nvii~Ly$\beta$ are enhanced, relative to their respective Ly$\alpha$ line intensities (Figure~\ref{fig:lyman}b).
However, for the CX process concerned here, the dominant capture channels are around the major quantum number $4\lesssim n\lesssim 6$ \citep[e.g.][]{smith12}.
Thus one would expect that the \oviii~Ly$\delta$ is stronger than the Ly$\gamma$ and that the \nvii~Ly$\gamma$ is stronger than the Ly$\beta$, which are, however, inconsistent with the data (Figure~\ref{fig:apecCX}).
Similarly, the APEC+ACX model overpredicts the enhancement of the \ovii~He$\gamma$ (17.768 \AA).

The model also fails to explain the iron line ratios. 
In CX, \fexvii\ and \fexviii\ lines come from Fe$^{17+}$ plus Fe$^{18+}$ ions, respectively.
However, in the temperature range of 0.2-0.3 keV, the ion fraction of Fe$^{17+}$ plus Fe$^{18+}$ is too small ($<5$\%) to produce any significant lines.

Last but not least, the interface area derived from the fitted CX model is not likely physically feasible.
The best-fit `norm' parameter of the ACX model means an ion incident rate of $\sim5^{+2}_{-1}\times10^{49}\,{\rm s^{-1}}$.
Taking the rotating velocity of the nuclear spiral ($\sim$200 \kmps) as the characteristic impact velocity between the hot and cold gas, and the median density of the hot gas as 0.05 cm$^{-3}$, we infer from the ion incident rate an interface area of about 5 kpc square.
This area is comparable to the cross section of the whole bulge region and is probably too large for the M31 nuclear spiral to generate.

\subsection{Resonant scattering (RS)}

RS can also dilute the {\it r} line intensities measured from a limited field \citep{chen18}.
A photon in a resonant line, such as \oviii~or \ovii~{\it r}, can be absorbed and then reemitted in a different direction. 
This directional redistribution of photons depends on their oscillator strength and can change the relative intensities of lines in a spectrum extracted from a field enclosing only part of the scattering medium.
In the case for the present study, the large \ovii~G-ratio observed toward the central region of the M31 bulge could, in principle, be due to RS of the \ovii~{\it r} flux. 
Indeed, a preliminary test of this scenario gives a reasonable explanation for the \ovii$+$\oviii\ complex (18-23 \AA) in the ``broad region'' RGS spectrum \citep{chen18}. 
This RS modeling gives an estimate of the optical depth up to $\sim8$ for the \ovii~{\it r} line along the line of sight to the galaxy center; about 50\% of the line flux could have been scattered out of the spectral extraction region.
The optical depths of \oviii~18.976 \AA\ and 18.973 \AA\ are $\sim4$ and $\sim2$ respectively, and about 30\% of the total \oviii~line flux could be out.

Here we examine the RS scenario based on some new spectroscopic diagnostics listed in Table~\ref{tab:evaluation}, as well as the spatial distribution of the diffuse X-ray emission observed with the \chandra\ images. 

Qualitatively, the scenario explains the different SB profiles in the 0.3--0.7 and 0.7--1.5 keV bands. 
In the latter, even the relatively strong resonant lines such as \neix~{\it r} and \fexvii~(15.014 \AA) have insufficiently large opacities at the temperature of $\sim0.2$ keV to significantly affect the spatial distribution of the emission. 
In contrast, the 0.3--0.7 keV band contains \ovii~{\it r} and \oviii~Ly$\alpha$ which suffer the RS spatial broadening effect most seriously, and other resonant lines such as \cvi~Ly$\alpha$, \nvii~Ly$\alpha$, and \nvi~{\it r}.
As a consequence, the 0.3--0.7 keV emission is more extended.
However, this explanation for the observed spatial distribution in the 0.3--0.7 keV band is somewhat problematic, because the required RS effect overproduces the broadening of the \oviii~line (Figure~\ref{fig:apec}).
One may need to attribute at least part of the spatial distribution to the decreasing plasma temperature with the off-center distance.

Then we look into the relevant spectroscopic diagnostics to further test the RS scenario.
Relatively, \oviii~Ly$\alpha$, as well as \ovii~{\it r}, have the highest RS opacities and are diminished in flux. 
Therefore, the observed high intensity-ratios of lines such as \oviii~Ly$\beta--\gamma$, and \nvii~Ly$\alpha$ to \oviii~Ly$\alpha$, can be explained, at least qualitatively. 
Because RS tends to redistribute the affected photons from the inner to the outer regions, one also expects that such ratios, as well as the \ovii~G-ratio, decrease from ``region a'' to ``c,'' as observed.
The spatial broadening of the \ovii~{\it r} line emission also accounts for part of the 21.8 \AA\ excess, alleviating it to a 2$\sigma$ confidence level \citep{chen18}.

\begin{figure}[htbp] 
\centering
      \includegraphics[angle=0,width=3.5in]{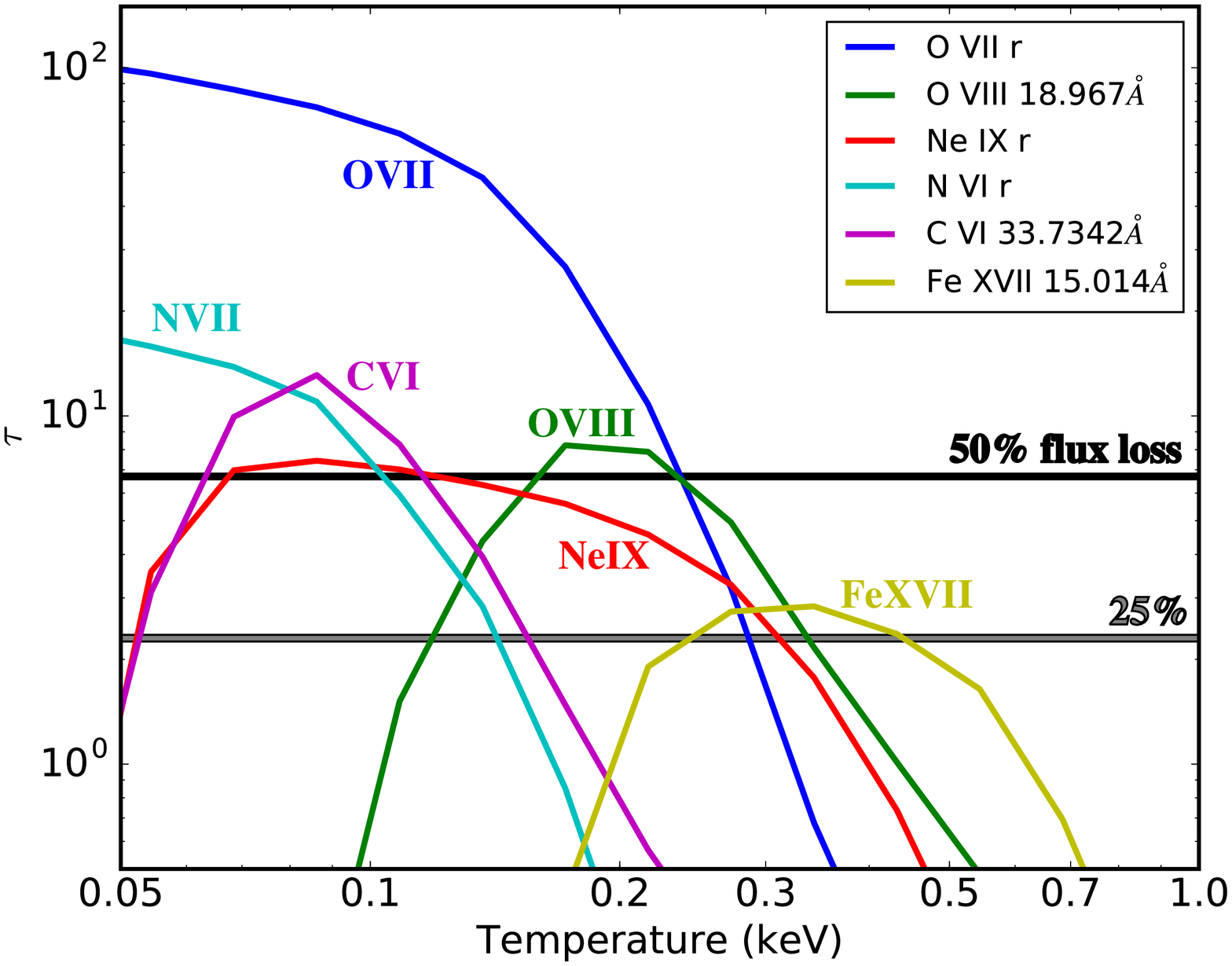} 
 \caption{The line opacity $\tau$ as a function of the plasma temperature for \ovii~{\it r}, \neix~{\it r}, \nvi~{\it r}, \fexvii~15.014 \AA, and \cvi~Ly$\alpha$ lines, assuming the solar abundances and an isotropic turbulence of $\sim$40 \kmps\ (from \citealt{chen18}). The gray and black horizontal lines correspond to the cases of losing 25\% and 50\% flux, respectively.}
\label{fig:tau} 
\end{figure}

For a more realistic comparison with the spectroscopic signatures obtained via the `broad' region spectrum, we present a detailed analysis of RS effects. 
Figure~\ref{fig:tau} shows the opacities ($\tau$) of six lines, which suffer RS mostly: \ovii~{\it r}, \oviii~(18.967 \AA), \neix~{\it r}, \nvi~{\it r}, \cvi~(33.7342 \AA), and \fexvii~(15.014 \AA), versus the temperature.
The calculation of these opacities assumes an isothermal hot plasma with a 3D spatial distribution described by a $\beta$-model and a turbulence velocity of 40 \kmps\ derived in \citet{chen18}, but using the solar abundances.
At the reference temperature of 0.2 keV that is preferred by the oxygen lines, for example, the RS effect can be sufficiently significant to explain the high G-ratio of \neix, but the opacities for \nvi~{\it r}, \cvi~(33.7342 \AA), and \fexvii~(15.014 \AA) are all smaller than one. 
So the RS scenario has difficulties in accounting for the large \nvi~G-ratio, the \oviii~Ly$\gamma$/Ly$\beta$ ratio, the \cvi~Ly$\zeta$/Ly$\gamma$ ratio, or the iron line ratios, as well as the increasing \nvii~Ly$\alpha$/\oviii~Ly$\alpha$ from the inner to the outer region, at least under the isothermal plasma assumption. 

Relaxing this assumption to allow for a temperature distribution would, of course, allow for more flexibilities. 
For example, if the temperature varies from $\sim 0.5$ to 0.1 keV from the inner to the outer region, we could have the corresponding RS opacity enhancements for \fexvii~(15.014 \AA), \nvi~{\it r}, and \cvi~lines simultaneously (Figure~\ref{fig:tau}). 
While the exploration of the full range of such flexibilities is difficult, if not impossible, we here focus on discussing spectroscopic signatures that are most difficult to be accounted for by the RS scenario alone. 

First, the expected RS effect on the \fexvii~(15.014 \AA) line is problematic.
We fit the line in the RGS spectrum with a Gaussian model, again allowing for a flexible width to account for the line profile broadening due to the spatial redistribution. 
Compared with the results from the previous narrow Gaussian fit (Table~\ref{tab:lines}), both the flux and the line profile change little, mainly because the blue side of the line is sharp and does not prefer a fat Gaussian.
There is no clear sign of the RS effect on this iron line.
If the RS effect is negligible, then the RS scenario cannot explain the considerably high \fexvii~15.261\AA/15.014 \AA\ ratio (without taking into account the \oviii~Ly$\gamma$).
On the other hand, let us ignore the above line profile argument and assume that  the \fexvii~15.014 \AA\ suffers significant RS.
In this case, the original flux of the \fexvii~15.014 \AA\  line would be higher, and it would significantly enhance the \fexvii~15 \AA/17 \AA\ ratio that has already suggested a far higher temperature $>2\times10^7$ K.

Second, the high \nvi~G-ratio is not likely explained by the RS, because the $\tau$ value of \nvi~{\it r} is always about one order of magnitude smaller than that of \ovii~{\it r} line over the entire temperature range (Figure~\ref{fig:tau}). 
For example, for the best-fit $\tau\sim8$ of \ovii~{\it r}, the $\tau$ of \nvi~{\it r} will be less than one under any temperature distribution, unless the abundance ratio of N to O is substantially higher than the assumed solar value.

\subsection{A brief summary}
In short, the plasma temperature in the M31 bulge likely decreases from the inner to the outer region, though the origin of this variation still needs to be explored. 
Allowing for the variation, the CX or RS scenario could reasonably explain many spectroscopic signatures listed in Table~\ref{tab:evaluation}.
However, significant difficulties remain. 
CX is insufficient to account for the line intensities of the high-order Lyman series and requires too large an interface area to be plausible in the M31 bulge.
RS still cannot explain the high G-ratio of \nvi~He$\alpha$.
Both scenarios cannot explain the high iron line ratios, the 21.8 \AA\ excess, and some of the spatial variations.
Therefore, although the CX and RS are expected in the region, they are not likely the dominant mechanism for the non-CIE spectroscopic signatures in the RGS spectra.

\section{AGN Relic Scenario}

An AGN comes and goes in a galaxy. 
The disappearing of an AGN itself, in particular, could be a sudden event, but its effect on the surrounding plasma, or an AGN relic, could last much longer. 
One example for such a relic is  the giant ionized nebula, discovered as Hanny's Voorwerp. 
This nebula represents a light echo of a powerful quasar ($L_{\rm bol}\sim10^{46}$ \ergs) less than 0.2 Myr ago at the center of the galaxy IC 2497\citep{lintott09}. 
Since then, more such AGN relics have been observed \citep{keel12, keel15, schirmer13}, and the disappearing of these AGN seems to occur mostly on time scales less than 0.1 Myr. 
Our Galaxy also likely hosted an energetic AGN more than 1 Myr ago, responsible for such phenomena as the Fermi bubbles \citep{guo12, su12} and the H$\alpha$ excess seen in the Magellanic Stream \citep{bland13}. 

Circumstantial evidence for a past AGN is present within the central several hundred pc of M31.
There is an apparent deficiency of molecular gas \citep[e.g.][]{melchior17, li19}, which might be understood as the AGN outburst having cleaned up preexisting gas in this region.
The dynamical timescale to rebuild the cold gas should be comparable to or longer than 1 Myr. 
On the gamma-ray perspective, though the results of Fermi bubble-like structures in M31 are rather inconclusive \citep[e.g.][]{li16, pshirkov16, feng19}, some past AGN activities may help explain the extended galactic center excess of gamma-rays that cannot be fully accounted for with typical astrophysical cosmic-ray sources such as supernovae or millisecond pulsars \citep[e.g.][]{mcdaniel19}. 
These old cosmic-ray electrons consequently enhance the radio synchrotron radiation within the central region through the strong magnetic field \citep{giebubel14}.

One may expect that the over-ionization effect of a past AGN can last even longer on surrounding diffuse hot plasma due to its low density. 
We find that a recombining AGN relic may provide a natural explanation for the spectral and spatial properties of the soft X-ray emission from the inner bulge of M31. 
In the following, we qualitatively compare various relevant X-ray spectroscopic diagnostics expected from this relic, based on a simple toy model.
A more detailed quantitative comparison with the data will be presented in a later paper.

\subsection{X-ray spectroscopic diagnostics of AGN relics}
Our toy model presented here is meant to capture the key characteristics of the X-ray emission from an AGN relic. 
The AGN was assumed to have lasted long enough to establish a photoionization equilibrium in the surrounding hot plasma and was shut off completely and abruptly. 
The initial condition of the plasma is calculated with Cloudy v13.11,\footnote{https://www.nublado.org} using the solar metallicities and the default setup for the AGN spectral energy distribution (SED) that is similar to a typical one \citep{elvis94}.
The ionization parameter $\xi=L_{\rm ion}/(n_{e}R^{2})$ is the input parameter, where $L_{\rm ion}$ is the ionizing luminosity, $n_{e}$ is the electron density, and $R$ is the distance to the AGN.
Cloudy returns the ion population for each assumed $\xi$.
Simultaneously, the temperature is also determined through thermal equilibrium, which monotonically increases with rising $\xi$.
When the AGN is ``off,'' all ions recombine over time toward a CIE state with this electron temperature. 
The cooling efficiency of the recombining plasma is usually several times less than that in a CIE plasma.
Because the plasma density is less than 0.1 \cmcu\ in the present case, the cooling timescale should be long enough, and the temperature decrease can be ignored currently.
The {\sl pyatomdb}\footnote{https://github.com/AtomDB/pyatomdb} procedure is used for calculating the time-evolving ion populations based on the AtomDB, and finally the line emissivities.

\begin{figure*}[htbp] 
 \centering
         \includegraphics[angle=0,width=2.3in,height=2.in]{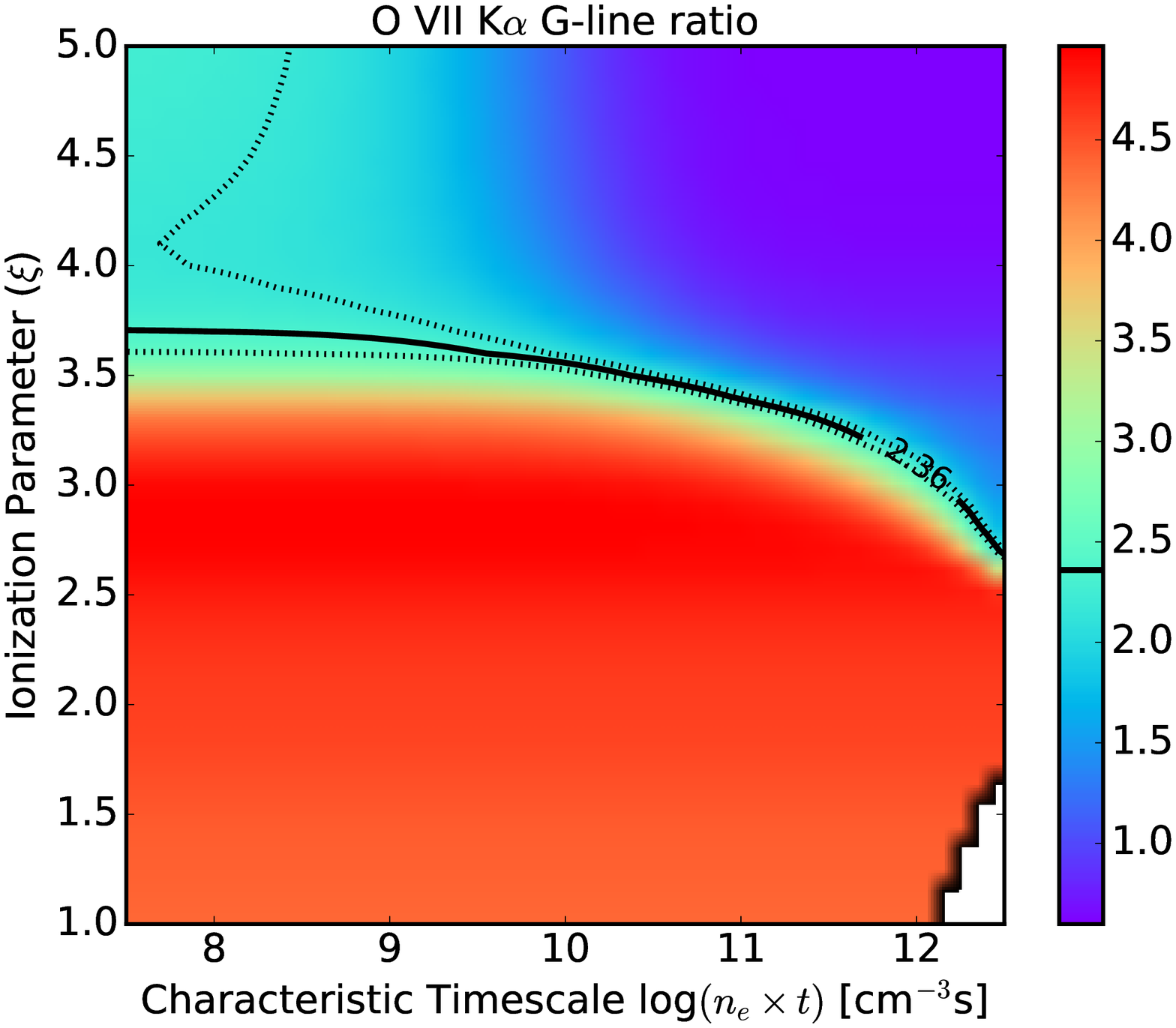}
         \includegraphics[angle=0,width=2.3in,height=2.in]{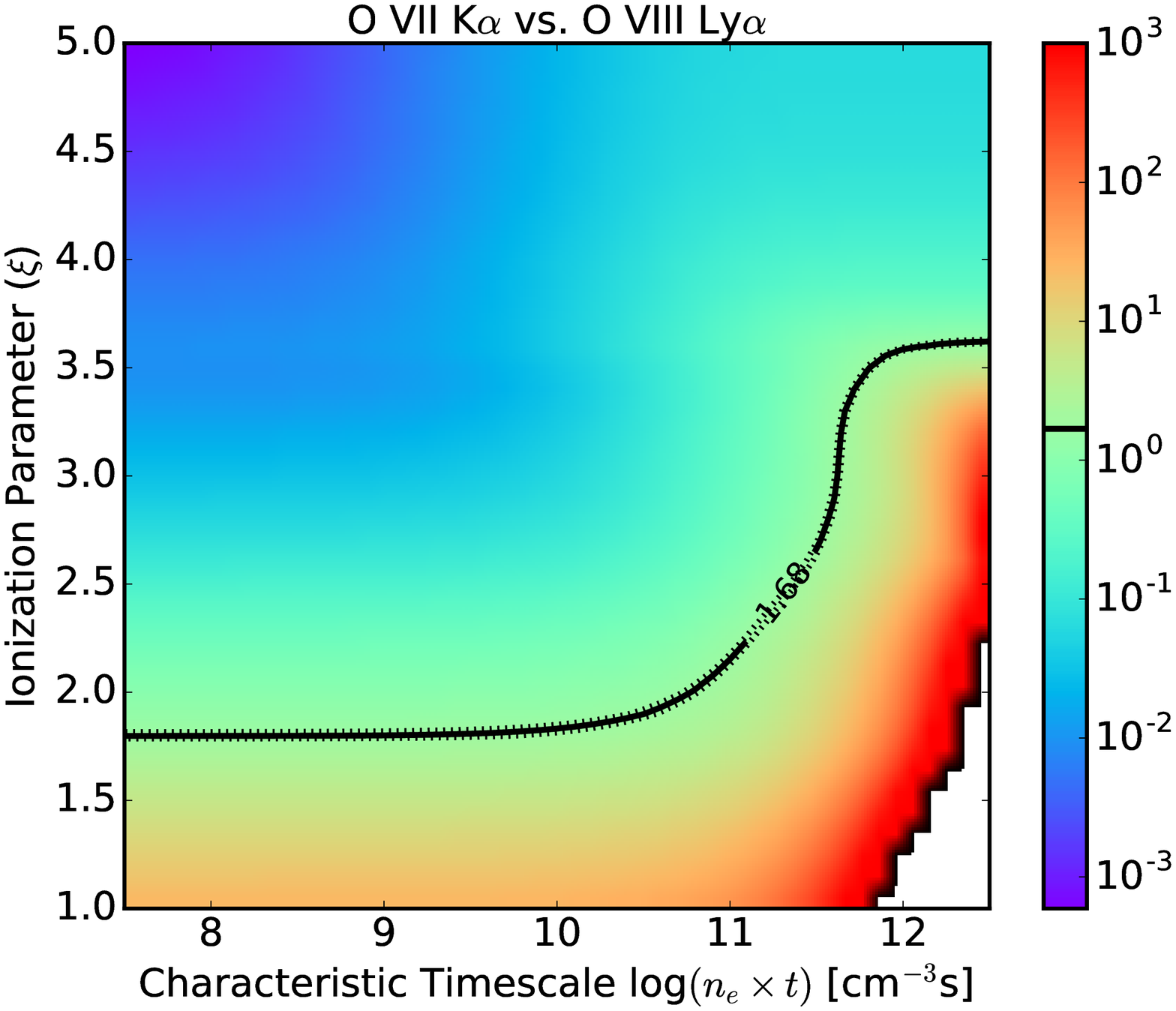}
         \includegraphics[angle=0,width=2.3in,height=2.in]{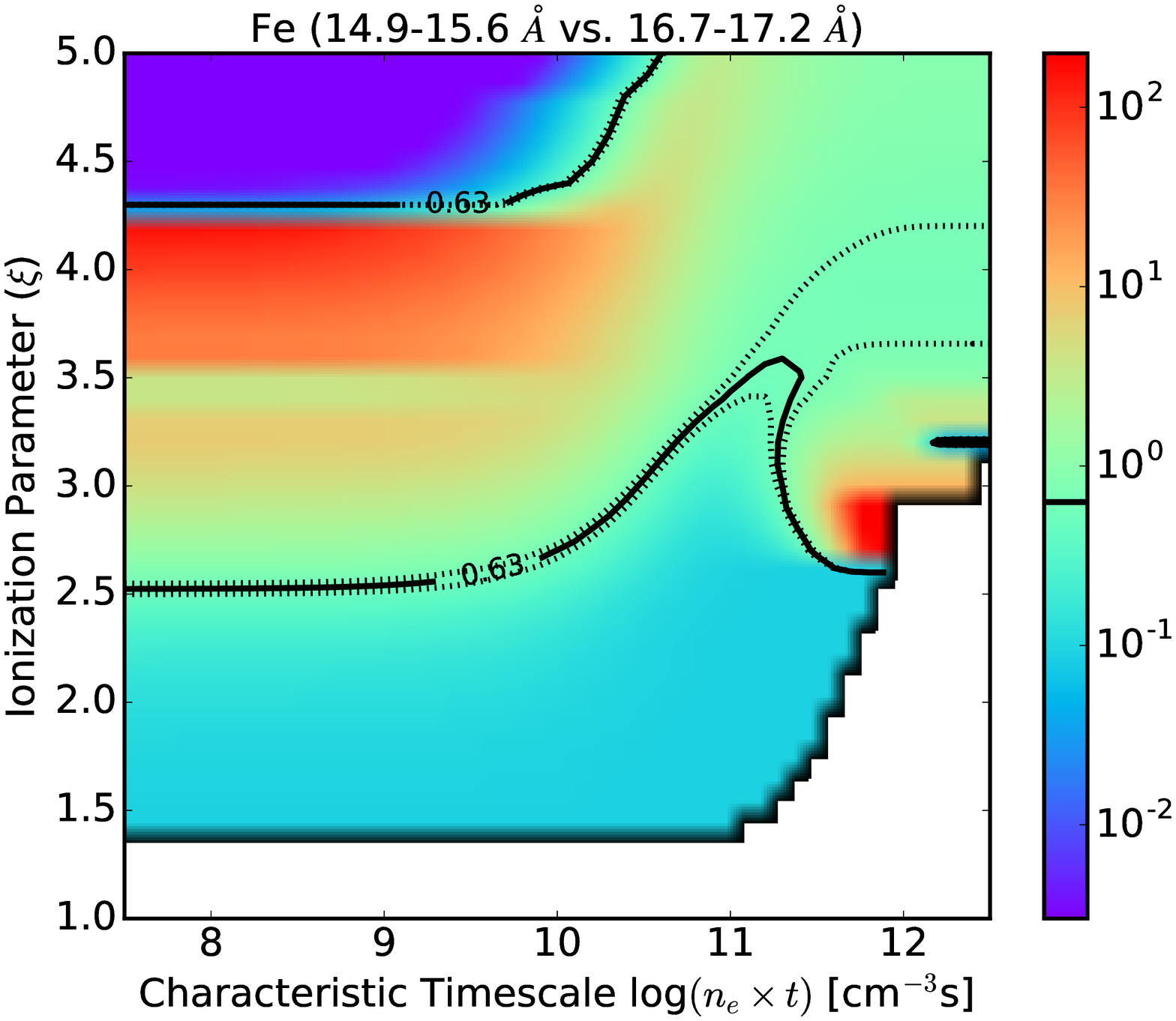}
         \includegraphics[angle=0,width=2.3in,height=2.in]{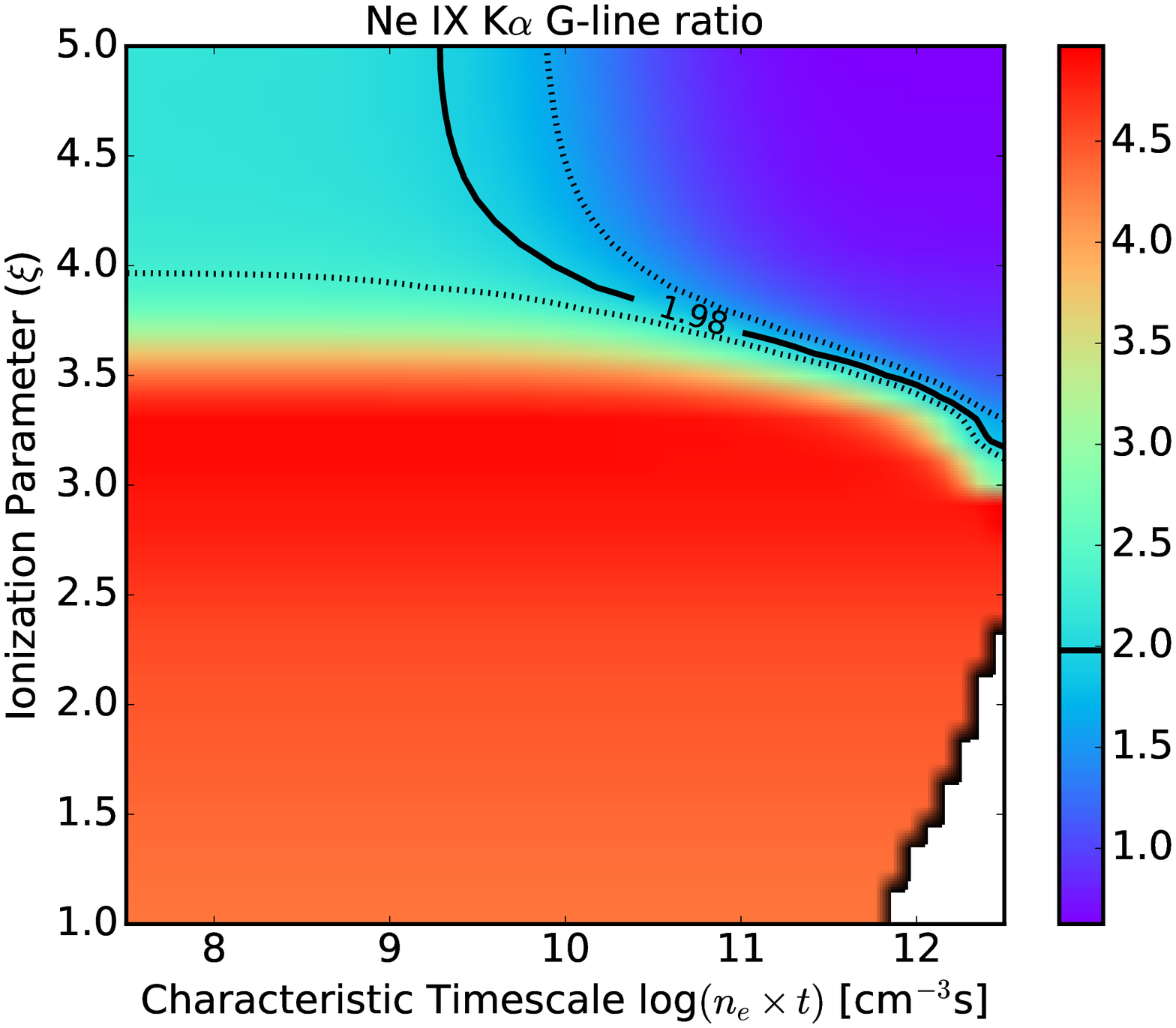}
         \includegraphics[angle=0,width=2.3in,height=2.in]{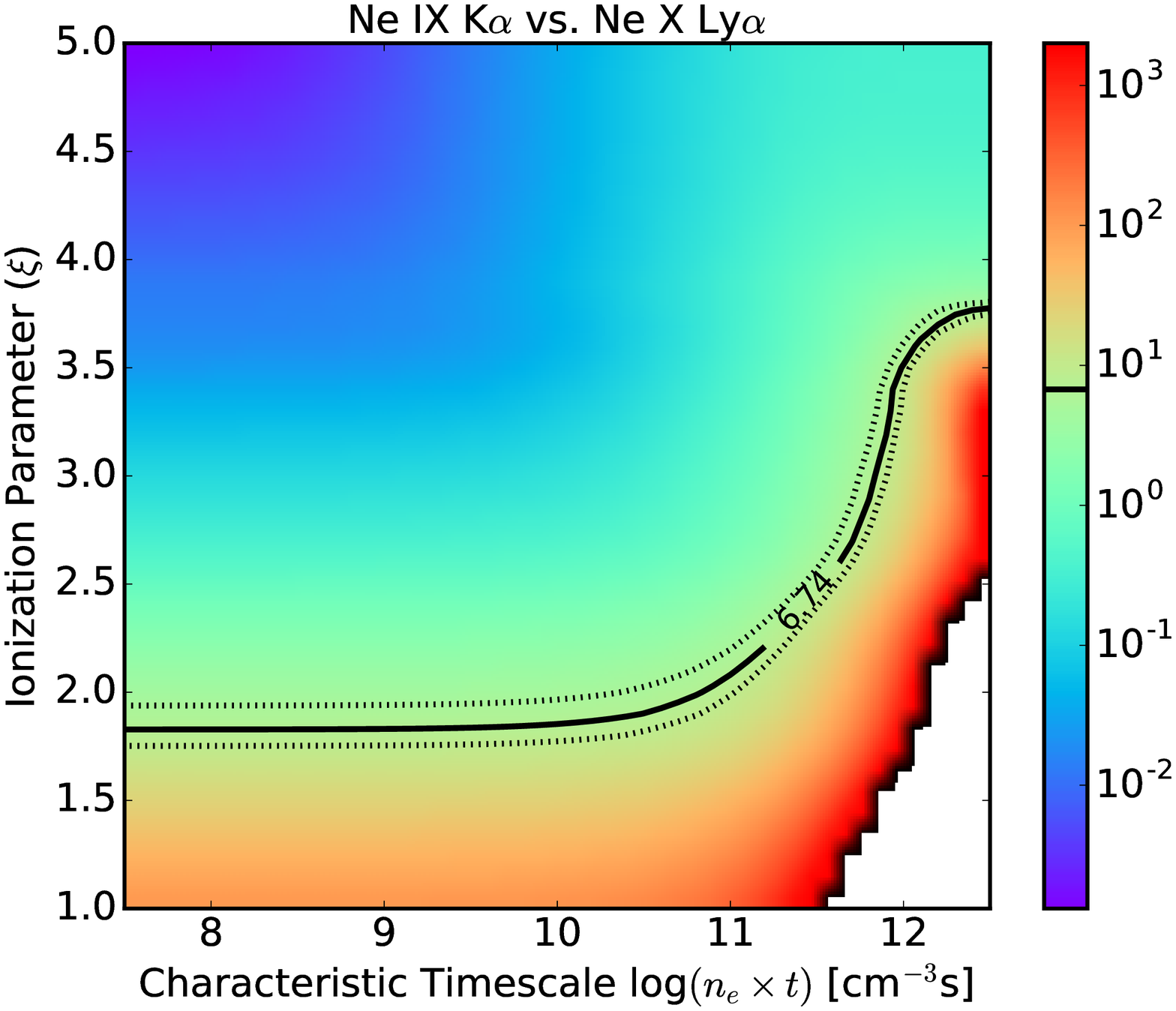}
         \includegraphics[angle=0,width=2.3in,height=2.in]{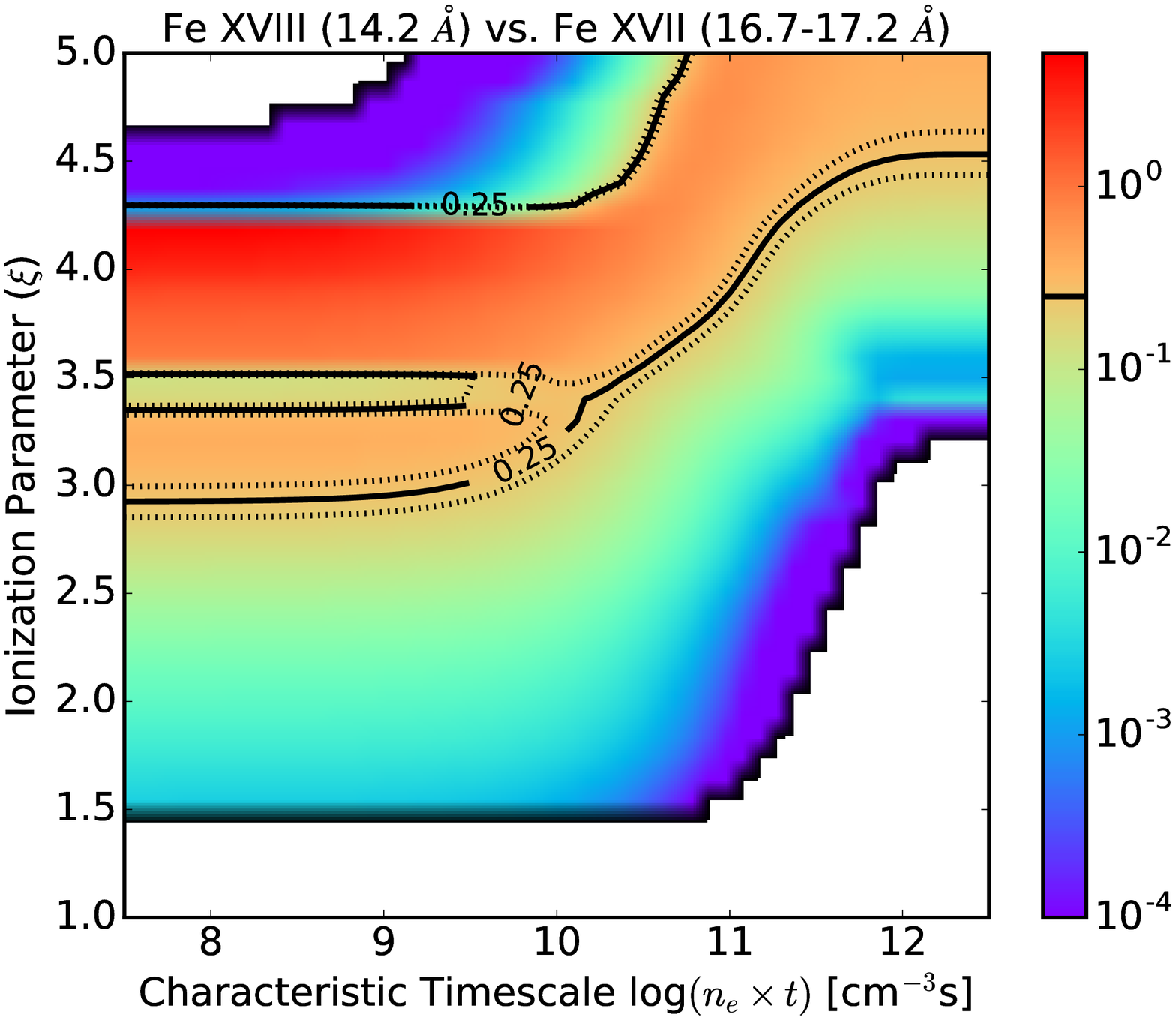}
         \hspace*{-0.3in}\includegraphics[angle=0,width=2.3in,height=2.in]{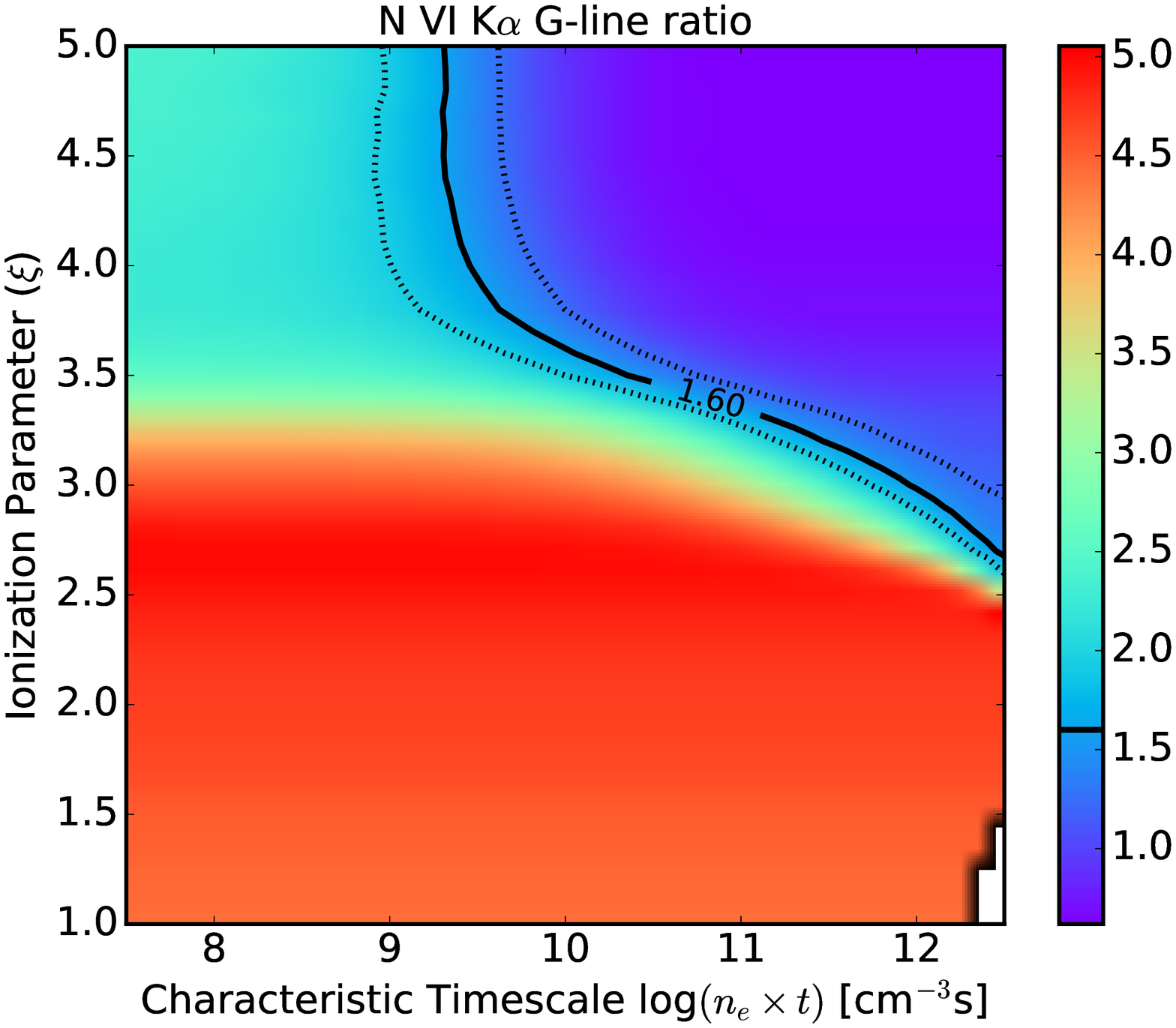}
         \includegraphics[angle=0,width=2.3in,height=2.in]{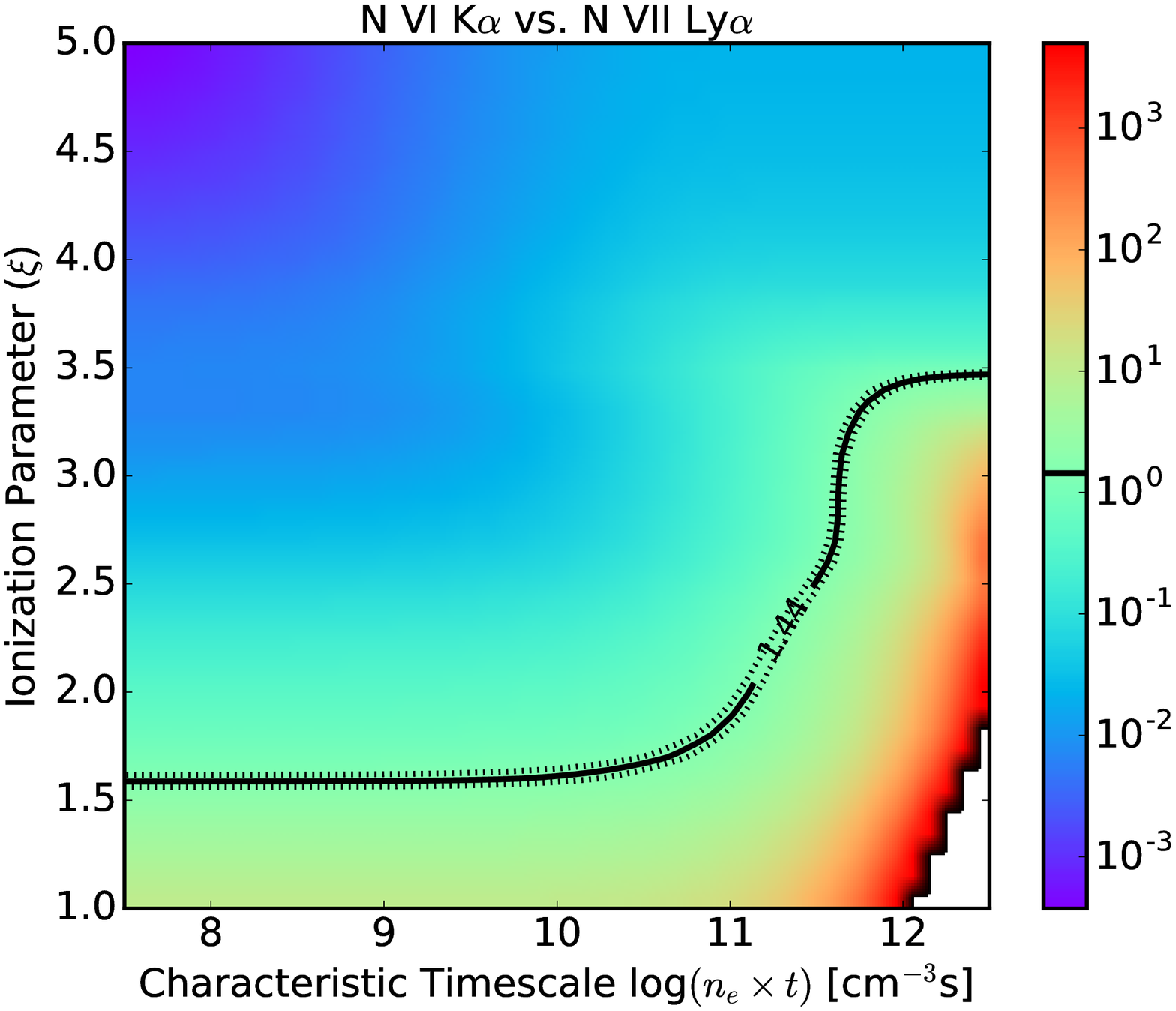}
         \includegraphics[angle=0,width=2.0in,height=2.in]{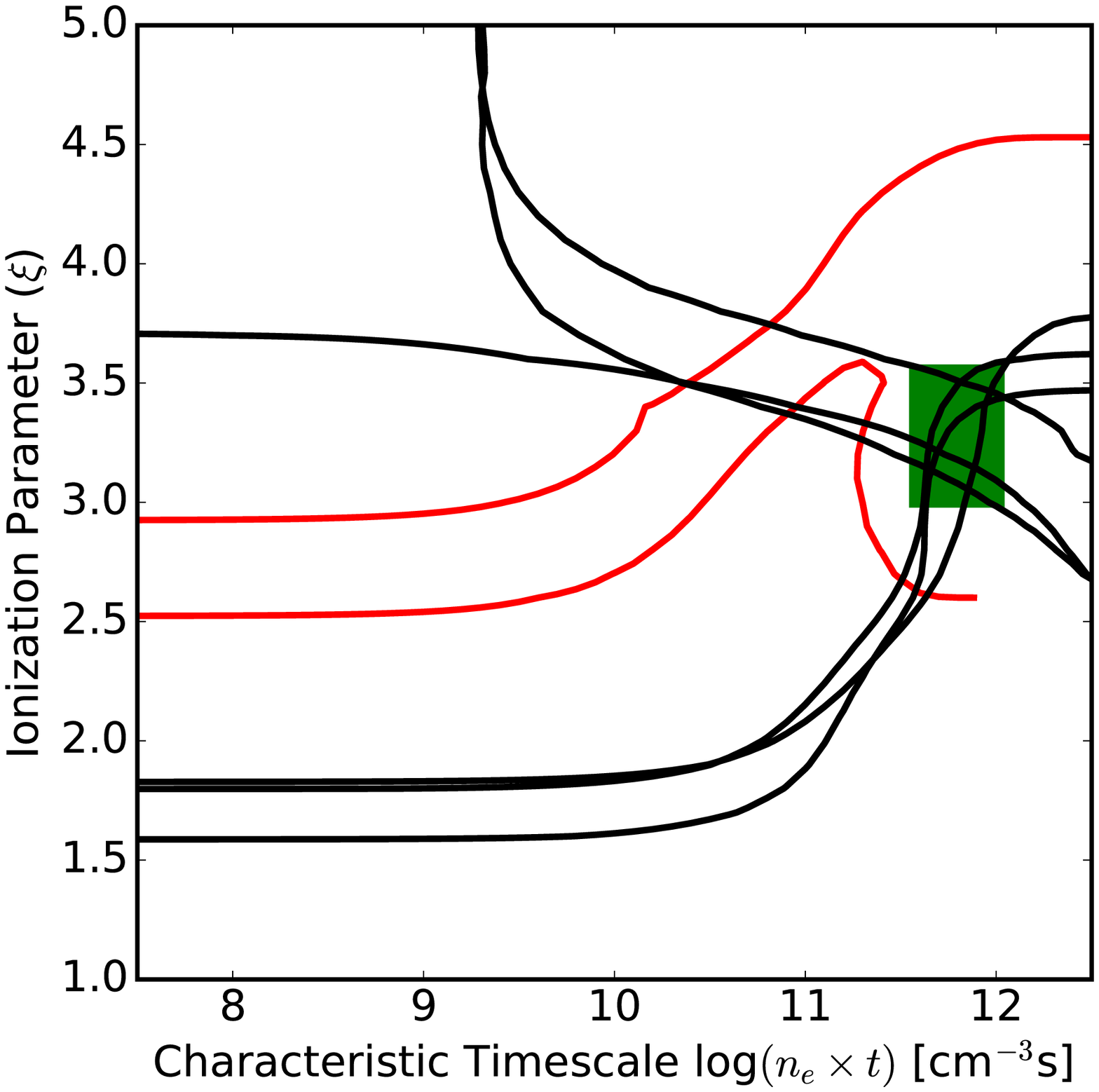}
         \caption{Key line ratios as the function of the ionization parameter ($\xi$) and the characteristic timescale ($n_e\times t$). The solid and dash contours are the line ratios and their corresponding 90\% confidence intervals measured from the ``broad region'' spectrum (see Table~\ref{tab:ratios}). The panel at the bottom-right corner summarizes the contour curves of line ratios, where the red curves are from iron, while the black curves are from other species. The green bar approximates the parameter space where these curves converge.}
 \label{fig:lines} 
\end{figure*}

Figure~\ref{fig:lines} presents various diagnostic line-ratio maps as a function of $\xi$ and characteristic ``timescales'' log$(n_e\times t $ (cm$^{-3}$s)). 
Overplotted on these line-ratio maps are the best-fit values and the 90\% uncertain ranges of the ratios measured from the ``broad region'' spectrum (Table~\ref{tab:ratios}). 

The parameters of $\xi$ and $n_e\times t$ constrained by the measurements converge well into a minimal region in the parameter space, as shown in the final panel of Figure~\ref{fig:lines}. 
The ionization parameter $\xi$ of the relic plasma is in the range of 3--4. 
The characteristic timescale is $\sim8 \times 10^{11}$ (cm$^{-3}$s), which corresponds to an AGN age of $\sim 5 \times 10^{5}$ yr, assuming a median density of 0.05 cm$^{-3}$ at the off-center distance of $\sim350$~pc, according to the $\beta$-model.

The apparent deviations of the measured iron lines from the minimal parameter range in the summary panel require some explanation. 
One possibility is that our measurement of the  \fexviii~(14.2 \AA) is an overestimation, because we have not considered the presence of the radiative recombination continuum (RRC) in the RGS spectra. 
In a recombining plasma, as considered here, \oviii~RRC (14.228 \AA) and \ovii~RRC (16.771 \AA) are likely significant, although they cannot be resolved out in the RGS spectra. 
These two RRC edges, convolved with the angular structure function of the 0.7--1.5 keV image, may have contaminated the \fexviii~(14.2 \AA) line, as well as the \fexvii~15.014 \AA\  line.  
The true \fexviii~(14.2 \AA)/\fexvii~(17 \AA) isopleth could be more consistent with the above constrained parameter range.   
  
That the AGN relic scenario explains all the line ratios above is physically understandable. 
One can see that the G-ratios of the He$\alpha$ triplets are generally high in a recombining plasma and decrease with time, as hydrogen-like ions are recombined into helium-like. 
Take it for an example.
Figure~\ref{fig:rrc}a shows the \ovii~{\it r} and {\it f} line emissivities from of O$^{5+}$, O$^{6+}$, and O$^{7+}$ ions as a function of the plasma temperature. 
Compared with a CIE case, an over-ionized plasma tends to have more O$^{7+}$ over a broad temperature range. 
The recombination and subsequent cascading down through the $\rm 1s^12s^1--1s^2$ transition results in stronger \ovii~{\it f} emission, and high-order Lyman series lines. 
Figure~\ref{fig:lyman}c shows the Lyman series line ratios from O$^{8+}$ and N$^{7+}$ ions, where the high Ly$\beta$/Ly$\alpha$ ratios match the measurements well. 
Although the enhancement of the \oviii~Ly$\gamma$/Ly$\alpha$ ratio seems not large enough to explain the measured values in the ``region a and b,'' the RRC contaminations may be responsible for the overestimations of the \fexvii~(15 \AA) and \oviii~Ly$\gamma$ lines.

\subsection{Spatial properties in the AGN relic}

\begin{figure}[htbp] 
 \centering
  \subfigure[]{
\begin{minipage}[t]{3.3in}
\centering
        \includegraphics[angle=0,width=3.3in]{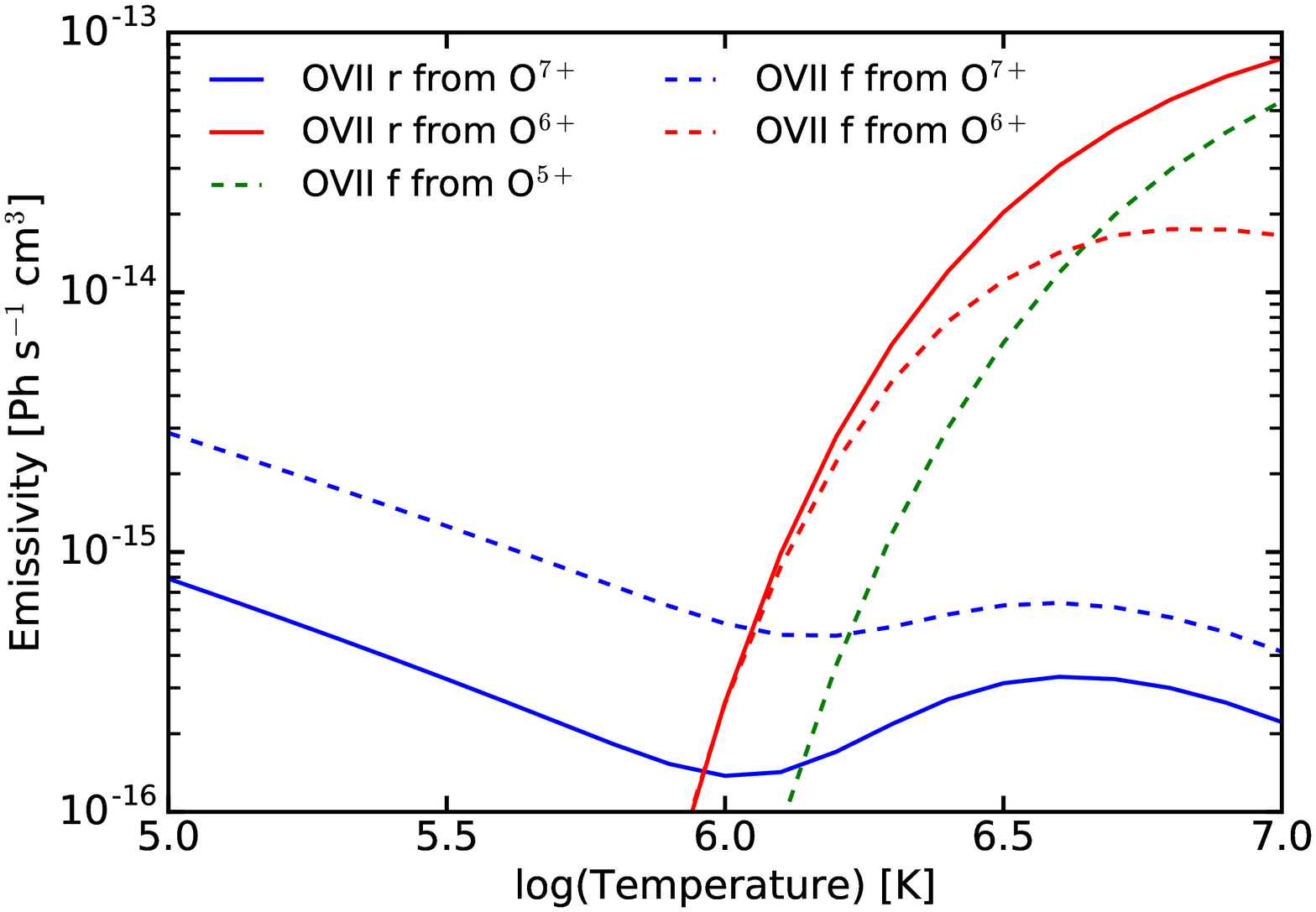}
\end{minipage}%
}%

 \subfigure[]{
\begin{minipage}[t]{3.3in}
\centering
       \includegraphics[angle=0,width=3.3in]{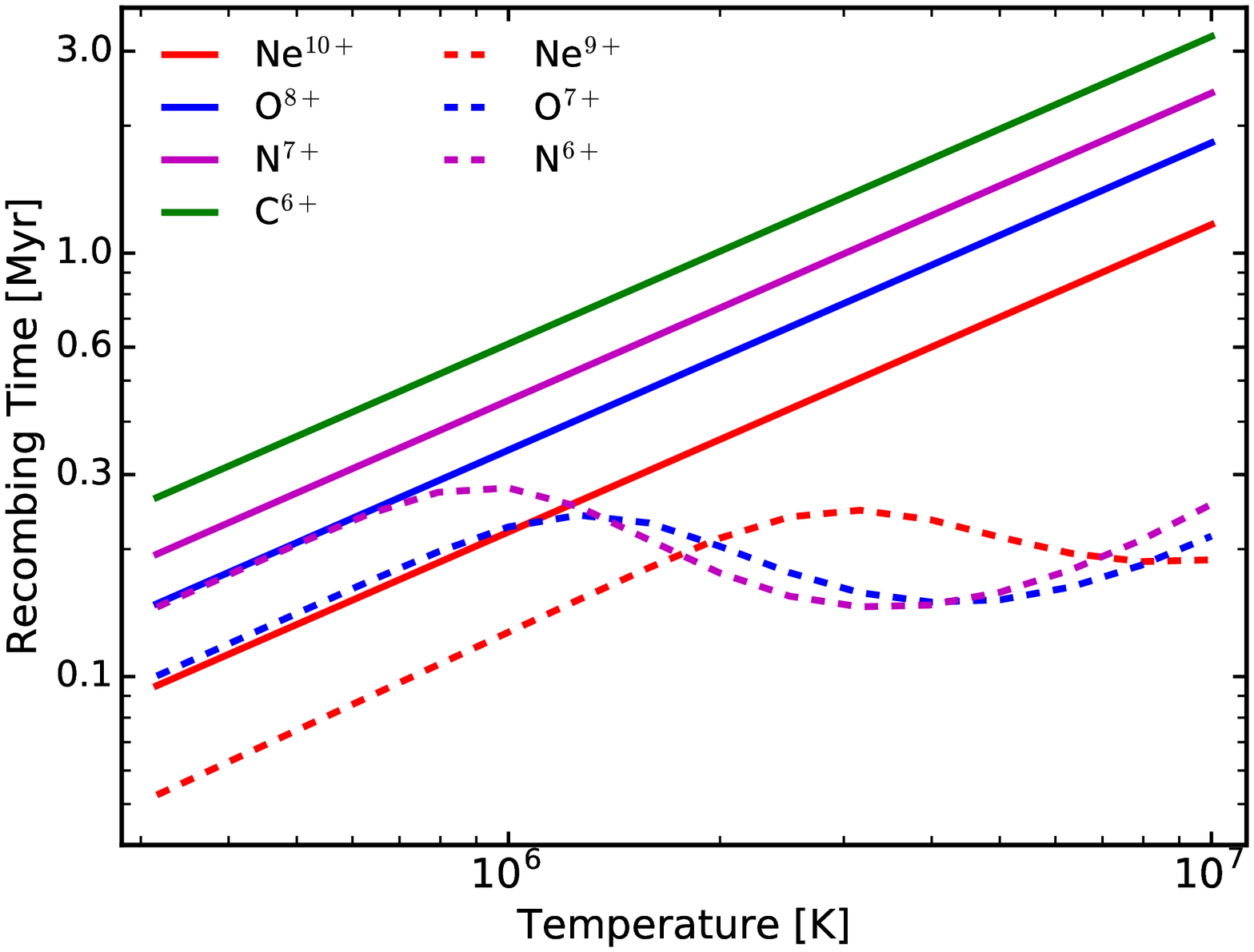}
\end{minipage}%
}%
\caption{Panel (a) shows the \ovii~{\it r} and {\it f} line emissivities from of O$^{5+}$, O$^{6+}$, and O$^{7+}$ ions as a function of temperature. The \ovii~{\it r} emissivity from O$^{5+}$ is negligible because remained electrons after the ionization can hardly stay at the higher energy level 1s2p 1P1. Panel (b) shows the recombining timescales for the fully ionized and H-like ions of C, N, O, and Ne, respectively, taking a density of 0.1 \cmcu\ for the illustration.}
\label{fig:rrc} 
\end{figure}

The above AGN relic scenario also offers a natural explanation of  the spatial variation of the spectroscopic signatures, as well as the apparent temperature decrease from the inner to outer parts of the bulge region. 
The outward temperature decrease is mainly due to the receding heating of the AGN before it was extinguished, though also affected by the differential cooling rate afterward.  
Except for the initial condition, the spectroscopic variation is further regulated by the differential recombining timescales of different ions.
Because the density variation in ``region a'' to ``c'' is within a factor of two, the timescale depends more on the temperature.
Figure~\ref{fig:rrc}b illustrates the recombining timescales for fully ionized and H-like ions, assuming the density $\sim 0.1$ \cmcu\ as an example. 
If the AGN was quenched 0.5 Myr ago, then the Ne$^{10+}$ ions at $3\times10^6$ K have the similar recombining rate as the C$^{6+}$ ions at $8\times10^5$ K.
At outer regions with lower temperature,  the \cvi~Lyman series lines can be prominent, while the Ne$^{10+}$ ions may be exhausted.
This trend explains the decrease of the Lyman series line ratios of \oviii~and \nex\ from the inner to the outer regions too.
The decrease of \fexviii~(14.208 \AA) from the ``region a'' to ``c,'' however, can be due to the decreasing \oviii~RRC contamination from the fully ionized ions.

Self-consistently, we should also consider the dynamics due to the AGN heating and the cooling of the relic. 
The heating should naturally lead to the expansion of plasma.
Considering the dynamics driven by thermal pressure as the first-order approximation, the hot gas at 100 pc could have been pushed outward for a few hundred pc over a half Myr, resulting in a lower-density cavity structure in the center. 
\citet{li09} also find an inward flattening in the SB within a radius of $\sim200$ pc in the \chandra\ image of M31.
A cavity-like structure of hot gas is also reported in the core of IC 2497, i.e., the associated galaxy of Hanny's Voorwerp \citep{sartori16}.

In this case, the significant \ovii~{\it r} emission may start at a shell-like structure, producing a double-peaked line in the RGS spectrum due to the spatial separation along the dispersion direction.
The \ovii~{\it f} would have this double peak feature for the same reason.
The left peak of \ovii~{\it f} superimposes on the right peak of \ovii~{\it r}, and together they could explain the excess around 21.8 \AA.
This kind of the spatial distribution effect due to the dynamical evolution of the AGN relic is expected to decrease from ``region a'' to ``c,'' because the double peaks gradually blend.

Last but not least, the bulk expansion should produce differential velocities of hot plasma at the different physical radius. 
As a result, RS effect can be significantly reduced, even without taking into account the turbulence induced by AGN outflows.

\section{Summary}
We have conducted a spatially resolved soft X-ray study of the M31 central bulge region.
This analysis is based chiefly on 766 ks of \xmm/RGS observations, complemented by 380 ks high spatial resolution data from \chandra~ACIS-S observations.
These deep observations provide unprecedented detailed spectral and spatial constraints on various scenarios that generate the soft X-ray emission from the region. 
The main results and conclusions of our analysis are as follows:

\begin{itemize}
\item We have identified multiple diagnostic X-ray spectroscopic signatures, which are not expected for an optically thin CIE plasma.
They include but are not limited to the large G-ratios of He$\alpha$ triplets (\ovii, \nvi, and \neix) and the high Lyman series line ratios (\oviii~Ly$\beta$/Ly$\alpha$ and Ly$\gamma$/Ly$\alpha$, and \nvii~Ly$\beta$/Ly$\alpha$).
The high iron line ratios (\fexviii~14.2 \AA/\fexvii~17 \AA\, and \fexvii~15 \AA/17 \AA) also suggest a temperature much higher than other line ratios.
We show that most of these signatures cannot be explained by such a CIE plasma, even allowing for temperature distribution.

\item The spectroscopic signatures further show significant spatial variation, which is also reflected in the radial surface brightness profiles of the diffuse emission in the 0.3--0.7 and 0.7--1.5 keV bands.
The profile in the softer band is considerably broader than that in the harder one. 
These variations indicate that the temperature of the plasma decreases with the increasing distance away from the galaxy's center.

\item  We show that part of the signatures (mostly the large G-ratios) could be explained by CX, as may be expected at the interface between the hot plasma and nuclear spiral arms of cold gas. 
However, this CX scenario has difficulties in explaining the Lyman series transition ratios and the iron line ratios. 
The scenario also demands too large an interface area that may not be plausible in the present relatively quiet  environment of the M31 bulge.

\item Similarly, RS could explain the higher G-ratios of \ovii~and \neix~triplets, and partly the energy-dependent surface brightness distribution, but is unable to account for the high \nvi~G-ratio and the high iron line ratios, even when the temperature variation is allowed. 
No significant RS spatial broadening effect, as may be expected for a plasma at a relatively high temperature ($\sim 0.3$~keV), is detected in the profile of the 15.014 \AA\ \fexvii~line.

\item We find that an AGN relic or a recombining plasma model provides a plausible explanation for essentially all the signatures, as well as the radial profiles, at least qualitatively. 
We infer from the model the presence of an AGN at the center of M31 about half a million years ago, and the initial ionization parameter $\xi$ of the relic plasma is in the range of 3--4. 
We argue that photoionization by AGN, which cycles on or off, can have profound effects on spectroscopic properties of diffuse hot plasma in and around galaxies, because of its long recombination timescale.
A more detailed and quantitative comparison between the AGN relic model and the RGS data will be presented in a later paper.

\end{itemize}

These results demonstrate that the spectra of the diffuse soft X-ray emission in M31 cannot be explained by equilibrium or simple CX or resonant scattering effects. 
Spatially resolved spectroscopic observations are critical to localizing the emission and understanding the underlying processes.

\section*{Acknowledgements}
We acknowledge the anonymous referee for careful reading and constructive comments. We thank Jiren Liu for the beneficial discussion. S.N. Zhang acknowledges the supports from NSFC grants 11573070 and 11203080, and the support of the China Scholarship Council. A. Foster acknowledges NASA APRA Grant 80NSSC18K0409.
Z.Y. Li acknowledges NSFC grant 11873028.

\label{lastpage}

\clearpage

\begin{appendix}
\section{A: Observation logs}
The Chandra/ACIS-S observations we utilized are listed in Table 5, and the XMM-Newton/RGS observations toward the bulge region of M31 are listed in Table 6.

\begin{deluxetable}{lcccccccr}[hb]
\tablecolumns{9}
\small
\tablewidth{0pt}
\tablecaption{\chandra/ACIS-S Observations\tablenotemark{*}}
\tablehead{\colhead{\chandra} &\colhead{R.A. (J2000)} & \colhead{Dec. (J2000)} & \colhead{Exposure} & \colhead{Roll Angle} & \colhead{OBS Start Date} & \colhead{OBS End Date} & \colhead{Data Mode} \\
\colhead{OBS. \# } & \colhead{(h~~m~~s)} & \colhead{($^\circ~~~^{\prime}~~~^{\prime\prime}$)} & \colhead{(s)} & \colhead{($^\circ$)} & \colhead{(yyyy-mm-dd)} & \colhead{(yyyy-mm-dd)}}
\startdata
13825    &00 42 43.607 & +41 16 14.51 &   1540 &    87.7 &2012-06-01 &2012-06-01 &FAINT \\
13825    &00 42 43.607 & +41 16 14.51 &  39788 &    87.7 &2012-06-01 &2012-06-01 &FAINT \\
13826    &00 42 43.571 & +41 16 13.83 &   1375 &    91.9 &2012-06-06 &2012-06-07 &FAINT \\
13826    &00 42 43.571 & +41 16 13.83 &  35824 &    91.9 &2012-06-06 &2012-06-07 &FAINT \\
13827    &00 42 43.533 & +41 16 13.11 &   1555 &    96.2 &2012-06-12 &2012-06-13 &FAINT \\
13827    &00 42 43.533 & +41 16 13.11 &  41478 &    96.2 &2012-06-12 &2012-06-13 &FAINT \\
13828    &00 42 43.467 & +41 16 11.05 &   1487 &   107.9 &2012-07-01 &2012-07-02 &FAINT \\
13828    &00 42 43.467 & +41 16 11.05 &  39878 &   107.9 &2012-07-01 &2012-07-02 &FAINT \\
14195    &00 42 43.448 & +41 16 06.45 &   1045 &   132.2 &2012-08-14 &2012-08-15 &FAINT \\
14195    &00 42 43.448 & +41 16 06.45 &  27680 &   132.2 &2012-08-14 &2012-08-15 &FAINT \\
14196    &00 42 44.556 & +41 15 57.54 &   1658 &   221.7 &2012-10-28 &2012-10-29 &FAINT \\
14196    &00 42 44.556 & +41 15 57.54 &  42644 &   221.7 &2012-10-28 &2012-10-29 &FAINT \\
14197    &00 42 43.497 & +41 16 04.36 &   1404 &   143.6 &2011-09-01 &2011-09-02 &FAINT \\
14197    &00 42 43.497 & +41 16 04.36 &  36634 &   143.6 &2011-09-01 &2011-09-02 &FAINT \\
14198    &00 42 43.524 & +41 16 03.67 &   1495 &   147.7 &2011-09-06 &2011-09-07 &FAINT \\
14198    &00 42 43.524 & +41 16 03.67 &  39872 &   147.7 &2011-09-06 &2011-09-07 &FAINT \\
15267    &00 42 43.451 & +41 16 06.46 &    416 &   132.2 &2012-08-16 &2012-08-17 &FAINT \\
15267    &00 42 43.451 & +41 16 06.46 &  11101 &   132.2 &2012-08-16 &2012-08-17 &FAINT \\
1575     &00 42 40.417 & +41 16 44.20 &  37308 &   180.4 &2001-10-05 &2001-10-05 &FAINT \\
1854     &00 42 41.637 & +41 15 19.73 &   4694 &   295.6 &2001-01-13 &2001-01-13 &FAINT \\
309      &00 42 45.120 & +41 16 43.57 &   5094 &    87.5 &2000-06-01 &2000-06-01 &FAINT \\
310      &00 42 43.947 & +41 16 43.53 &   5078 &   108.7 &2000-07-02 &2000-07-03 &FAINT \\
\enddata
\tablenotetext{*}{The exposure represents the live time (dead time corrected) of cleaned data.}
\label{tab:chandralog}
\end{deluxetable}

\begin{deluxetable}{lccccc}
\tablecolumns{6}
\small
\tablewidth{0pt}
\tablecaption{36 \xmm/RGS observations (2000-2012)}
\tablehead{\colhead{ID} & \colhead{RA} & \colhead{Dec} &  \colhead{P.A.} & \colhead{$t_{Exp}$} & \colhead{$t_{eff}$} }
\startdata
0109270101 & 10.680324 & 41.266006 & 76.1 & 57.9 & 33.5  \\ 
0109270501 & 10.680122 & 41.266124 & 76.1 & 10.6 &4.7  \\ 
0112570101 & 10.679958 & 41.259701 & 249.9 & 64.3 & 62.8  \\ 
0112570401 & 10.679071 & 41.266209 & 78.3 & 46.0 & 33.9  \\ 
0112570601 & 10.680028 & 41.260037 & 257.0 & 13.3 & 13.0  \\ 
0112570701 & 10.680006 & 41.259403 & 249.9 & 4.5 & 3.3  \\ 
0405320501 & 10.685502 & 41.272990 & 71.7 & 21.9 & 20.3  \\ 
0405320601 & 10.684280 & 41.273203 & 51.3 & 21.9 & 19.1  \\ 
0405320701 & 10.686068 & 41.265342 & 252.4 & 15.9 & 15.9  \\ 
0405320801 & 10.685275 & 41.265725 & 243.2 & 13.9 & 13.9  \\ 
0405320901 & 10.685723 & 41.266047 & 231.8 & 16.9 & 16.9  \\ 
0505720201 & 10.685294 & 41.265129 & 253.8 & 27.5 & 27.5  \\ 
0505720301 & 10.684693 & 41.266217 & 247.9 & 27.2 & 27.1  \\ 
0505720401 & 10.685944 & 41.265534 & 242.2 & 22.8 & 22.6  \\ 
0505720501 & 10.685340 & 41.266297 & 236.6 & 21.8 & 20.5  \\ 
0505720601 & 10.686532 & 41.265788 & 230.8 & 21.9 & 21.9  \\ 
0551690201 & 10.685436 & 41.266376 & 252.7 & 21.9 & 21.8  \\ 
0551690301 & 10.684949 & 41.265549 & 246.9 & 21.9 & 21.7  \\ 
0551690401 & 10.685264 & 41.266371 & 243.5 & 27.1 & 9.4  \\ 
0551690501 & 10.686075 & 41.266356 & 236.8 & 21.9 & 21.3  \\ 
0551690601 & 10.685709 & 41.265679 & 232.1 & 26.9 & 19.2  \\ 
0600660201 & 10.684718 & 41.265592 & 254.1 & 18.8 & 18.7  \\ 
0600660301 & 10.685422 & 41.265350 & 248.2 & 17.3 & 17.3  \\ 
0600660401 & 10.685428 & 41.266005 & 243.7 & 17.2 & 17.2  \\ 
0600660501 & 10.685658 & 41.266166 & 238.1 & 19.7 & 19.5  \\ 
0600660601 & 10.686281 & 41.265839 & 233.5 & 17.3 & 17.3  \\ 
0650560201 & 10.684482 & 41.265614 & 255.5 & 26.9 & 26.9  \\ 
0650560301 & 10.685331 & 41.265653 & 249.5 & 33.4 & 33.3  \\ 
0650560401 & 10.685800 & 41.265618 & 243.8 & 24.3 & 22.1  \\ 
0650560501 & 10.685515 & 41.266252 & 238.3 & 23.9 & 23.9  \\ 
0650560601 & 10.685947 & 41.265886 & 232.5 & 23.9 & 23.8  \\ 
0674210201 & 10.684554 & 41.266374 & 254.5 & 20.9 & 20.8  \\ 
0674210301 & 10.684424 & 41.266428 & 248.5 & 17.3 & 17.3  \\ 
0674210401 & 10.684782 & 41.266029 & 244.0 &19.9  & 19.9  \\ 
0674210501 & 10.685569 & 41.265826 & 240.7 & 17.3 & 17.3  \\ 
0674210601 & 10.684913 & 41.265902 & 235.0 & 26.0 & 20.4  \\ 
\enddata
\tablenotetext{*}{Columns are: Observation ID, RA and Dec, Position Angle (degree), Exposure Time (ks), and Effective Time (ks).}
\label{tab:log}
\end{deluxetable}

\clearpage

\section{B. Key lines in the RGS spectra}
Here we present the line identifications and flux measurements in detail for $\sim$ 30 key lines of Ne, O, N, C, and Fe, in the RGS spectra of M31.
These lines are fitted simultaneously using narrow Gaussians ($\sigma\sim0.001$ \AA) as described in Sec. 3.2.

The neon lines considered in the fit are \nex~Ly$\alpha$ and Ly$\beta$, and \neix~He$\alpha$ triplet.
The \nex~Ly$\alpha$ (12.134 \AA) blends with two \fexvii~lines (12.124 \AA\ and 12.266 \AA) that have roughly equal fluxes.
But now that the separation between the \nex~Ly$\alpha$ and \fexvii~12.266 \AA\ is 0.132 \AA, larger than the RGS spectral resolution of 0.07 \AA, their contributions can be reasonably constrained by jointly fitting.
The \nex~Ly$\beta$, though included in Table~\ref{tab:lines}, appears marginal in the ``region a'' spectrum.
The \neix~He$\alpha$ triplet, in principle, could be contaminated by \fexix~lines around 13.5 \AA.
However, this would require a temperature of hot gas above 0.7 keV; thus, these potential lines are not included in the fit.
We fix the flux of {\it i} line in \neix~triplet as 1/3.5 of the {\it f} line flux, where the ratio is nearly constant for a CIE plasma in the temperature range of $10^6-2\times10^7$ K.
The \neix~He$\beta$ line (11.544 \AA) is only covered by the RGS2 and unfortunately falls into a gap due to bad pixels on the chip. Thus it is not included.

The oxygen lines considered are \oviii~Ly$\alpha$--Ly$\epsilon$, and \ovii~He$\alpha$ triplet and He$\beta$.
The \oviii~Ly$\alpha$ blends with the \ovii~He$\beta$, which can be reasonably separated in a fit.
However, the \oviii~Ly$\beta$ could be contaminated by the presence of two \fexviii~lines (16.004 \AA\ and 16.071 \AA, the latter of which has twice the flux of the former), when the temperature is larger than 0.4 keV in a CIE plasma.
We fit this \oviii~Ly$\beta$ complex with a single Gaussian in the ``broad region'' spectrum, with its center allowed to vary.
This fit gives a central wavelength of 15.98$\pm$0.02 \AA, which matches the redshifted wavelength (15.99 \AA) of \oviii~Ly$\beta$ well, and hence indicates that the \fexviii~lines are negligible.
The \oviii~Ly$\gamma$ and Ly$\delta$ (14.821 \AA) are strongly blended with the \fexvii~lines (15.014 and 15.261 \AA).
We only infer the flux of \oviii~Ly$\gamma$ from the two \fexvii~lines. 
The \oviii~Ly$\epsilon$ (13.634 \AA) is marginally detected (e.g., in Figure~\ref{fig:gauss}), but is not included in the fit.
The \ovii~He$\alpha$ does not mix with other lines, but the three individual lines of the triplet are deeply blended.
We fix the {\it i} line flux as 1/4.44 of the {\it f} line value that this ratio is nearly constant in a CIE plasma, at least.

The nitrogen lines considered are \nvii~Ly$\alpha$--Ly$\delta$, and \nvi~He$\alpha$ triplet.
We do not include \nvi~He$\beta$ (24.898 \AA) because of its relative weakness and wavelength closeness to the strong \nvii~Ly$\alpha$.
The \nvii~Ly$\gamma$ and Ly$\delta$ blend with the \oviii~Ly$\alpha$ and are fitted according to their wavelengths.
The \nvi~He$\alpha$ triplet has a better separation of {\it r} and {\it f} lines than the \ovii~He$\alpha$.
However, the \nvi~{\it r} line is slightly blended with the \cvi~Ly$\beta$.
We fix the \nvi~{\it i} line flux as 1/6.85 of the {\it f} line flux.

The carbon lines considered are \cvi~Ly$\alpha$--Ly$\zeta$.
The \cvi~Ly$\alpha$ is broader than other lines and prefers the broadening profile from the 0.3--0.7 keV image, but we still fit it based on the profile from the 0.7--1.5 keV image for the consistency.
The \cvi~Ly$\beta$ slightly blends with the \nvi~{\it r} line, while the Ly$\gamma$ is more isolated.
The \cvi~Ly$\delta$--Ly$\zeta$ are close in wavelength, and their measurements are affected by the red wing of the strong \nvii~Ly$\alpha$.

The iron lines considered are \fexviii~at 14.208 \AA\ and other three groups of \fexvii~lines, the most prominent lines among a bunch of inner shell transitions of iron.
The first group is \fexvii~lines at 12.124 and 12.266 \AA, the latter of which has a flux of $\sim$92\% of the former in a CIE plasma.
The two lines are mixed with the \nex~Ly$\alpha$, but their contribution can be disentangled. 
The second group is three \fexvii~lines around 17 \AA, which are not blended with other lines.
We fit them individually but mostly treat them as a unity in the analysis.
The third group is \fexvii~lines at 15.014 and 15.261 \AA.
In a CIE plasma, the flux of the latter is about one-third of the former.
However, if we fit the two lines with a fixed ratio, a prominent residual around 15.2 \AA\ would pop out in all the spectra.
The fit can be even worse if two \fexvi~lines (14.955 and 15.050 \AA) are taken into account.
A possible contributor is \oviii~Ly$\gamma$ (15.176 \AA).
We do not try to fit the \oviii~Ly$\gamma$ plus Ly$\delta$ together with the two \fexvii~lines, because the four lines are too close in wavelength to be separated.
Instead, we fit the two \fexvii~lines only and then infer the flux of \oviii~Ly$\gamma$:  \fexvii~[15.261 \AA]$-$(\fexvii~[15.014 \AA]/3).
In this way, the flux of \fexvii~15.014 \AA\ is maximized.
The \fexviii~14.028 \AA\ seems distinctive in the spectra of all regions, though the presence of this line requires a temperature higher than 0.4 keV.
Meanwhile, in a CIE plasma, the summed flux of \fexviii~lines at 16.004 and 16.071 \AA\ is about 80\% of the flux of the \fexviii~14.028 \AA.
However, in the ``broad region'' spectrum the center of the 16 \AA\ feature matches the \oviii~Ly$\beta$ well and suggests that the two \fexviii~lines at 16 \AA\ are negligible, as discussed for oxygen lines.
A test is also done for fitting the two \fexviii~lines (with a fixed ratio of 0.6) together with the \oviii~Ly$\beta$ in the ``region a'' spectrum, where the line broadening profile is the narrowest. 
The result gives a small upper limit to the flux of \fexviii~16.071 \AA\ as $10^{-6}\,{\rm ph\,s^{-1}\,cm^{-2}}$, much smaller than the \fexviii~14.208 \AA\ flux.
As a result, we ignore the two lines during the fitting.

\end{appendix}

\end{document}